\documentclass[twocolumn]{aastex62}
\usepackage{graphicx}
\usepackage[caption=false]{subfig}
\usepackage{natbib}
\usepackage{placeins}
\usepackage{amsmath}
\usepackage{tgcursor}
\usepackage[caption=false]{subfig}
\usepackage{booktabs}
\setcitestyle{notesep={ }}

\def\apjl{ApJ}                
\def\apjs{ApJS}               
\def\ao{Appl.~Opt.}           
\def\aap{A\&A}                

\def\apjl{Astrophys. J. Lett.}             		

\def\apjs{Astrophys. J., Suppl. Ser.}            
\def\aap{Astron. Astrophys.}               		

\begin{document}

\title{Organic complexity in protostellar disk candidates}

\author{Jennifer B. Bergner}
\affiliation{Harvard University Department of Chemistry and Chemical Biology, Cambridge, MA 02138, USA}
\email{jennifer.bergner@cfa.harvard.edu}

\author{Rafael Mart\'in-Dom\'enech}
\affiliation{Harvard-Smithsonian Center for Astrophysics, Cambridge, MA 02138, USA}

\author{Karin I. \"Oberg}
\affiliation{Harvard-Smithsonian Center for Astrophysics, Cambridge, MA 02138, USA}

\author{Jes K. J{\o}rgensen}
\affiliation{Niels Bohr Institute \& Centre for Star and Planet Formation, University of Copenhagen,
{\O}ster Voldgade 5-7, 1350 K{\o}benhavn, Denmark}

\author{Elizabeth Artur de la Villarmois}
\affiliation{Niels Bohr Institute \& Centre for Star and Planet Formation, University of Copenhagen,
{\O}ster Voldgade 5-7, 1350 K{\o}benhavn, Denmark}

\author{Christian Brinch}
\affiliation{Research Group for Genomic Epidemiology, National Food Institute, Technical University of Denmark, 2800 Kgs. Lyngby, Denmark}

\begin{abstract}
\noindent We present ALMA observations of organic molecules towards five low-mass Class 0/I protostellar disk candidates in the Serpens cluster.  Three sources (Ser-emb 1, Ser-emb 8, and Ser-emb 17) present emission of CH$_3$OH as well as CH$_3$OCH$_3$, CH$_3$OCHO, and CH$_2$CO,  while NH$_2$CHO is detected in just Ser-emb 8 and Ser-emb 17.  Detecting hot corino-type chemistry in three of five sources represents a high occurrence rate given the relative sparsity of these sources in the literature, and this suggests a possible link between protostellar disk formation and hot corino formation.  For sources with CH$_3$OH detections, we derive column densities of 10$^{17}$--10$^{18}$ cm$^{-2}$ and rotational temperatures of $\sim$200--250 K.  The CH$_3$OH-normalized column density ratios of large, oxygen-bearing COMs in the Serpens sources and other hot corinos span two orders of magnitude, demonstrating a high degree of chemical diversity at the hot corino stage.  Resolved observations of a larger sample of objects are needed to understand the origins of chemical diversity in hot corinos, and the relationship between different protostellar structural elements on disk-forming scales.
\end{abstract}


\section{Introduction}
Planet formation takes place in the gas- and dust-rich disks orbiting young stars.  The chemical inventories in these protoplanetary disks therefore influence the compositions of nascent planets.  It is of particular interest to origins of life studies to understand the chemistry of complex (6+ atom, hydrogen-rich) organic molecules (COMs) in planet-forming regions, since these species are considered to be precursors for prebiotic chemistry \citep{Herbst2009, Jorgensen2012}.

The process of star (and eventually planet) formation begins in the dense cores of molecular clouds.  The youngest protostars, termed Class 0, are still deeply embedded in their natal envelope.  Class I protostars are undergoing envelope infall and accretion onto a circumstellar disk, and Class II protostars have cleared their envelope and host Keplerian disks that feed accretion onto the star \citep{Lada1987, Andre1993, Dunham2014}.  Planet formation was historically thought to take place during the Class II stage.  In recent years, the high sensitivity and spatial resolution of ALMA has enabled the characterization of complex organic molecule emission in Class II disks, enhancing our understanding of organic chemistry at this stage \citep{Oberg2015, Walsh2016, Bergner2018, Loomis2018, Favre2018}.  However, there is growing evidence suggesting that planet formation begins at earlier evolutionary stages \citep{Harsono2018}.  Notably, high-resolution (sub-)mm continuum observations are revealing that dust sub-structure appears ubiquitous in Class II disks \citep{Andrews2018}.  One compelling explanation for this sub-structure is interactions of large (Neptune- to Jupiter-mass) planets with the disk \citep{Zhang2018}, which would require that the planet formation process begins in younger (Class 0/I) disks.  This scenario is supported by observations of sub-structure in the embedded disk HL Tau \citep{ALMA2015}, implying grain growth or even planet formation at an early evolutionary stage.

Some deeply embedded protostars are seen to host a rich and warm gas-phase organic chemistry on small ($\sim$100 AU) scales.  These sources, termed hot corinos, form when temperatures around the protostar exceed the water ice sublimation point, and ice mantles are liberated into the gas phase \citep{Herbst2009}.  At present, the physical nature of hot corinos is unknown because they are generally unresolved or marginally resolved in observations.  Several structural elements, including protostellar disks, centrifugal barriers, and outflows, occur on similar spatial scales within the protostellar inner envelope \citep{Sakai2014, Lee2014, Harsono2014, Yen2015}; without high-sensitivity and high-resolution observations it is difficult to disentangle if these elements are related or distinct in their physics and chemistry, and which, if any, are sources of hot corino chemistry.  

Since protostellar disks are likely the sites where planet formation begins, it is of great interest to understand whether the organic complexity detected in hot corinos is related to the material incorporated into the disk.  To date there is no conclusive tie between hot corinos and protostellar disks.  Some sources with hot corinos appear to show velocity structure in their molecular line emission consistent with the presence of a rotating disk \citep{Choi2010, Lee2014, Codella2014, Oya2016, Oya2017}.  Still, other hot corino sources show no clear signatures of a rotating disk \citep{Maury2014, Imai2016, Jacobsen2018}.  More observations are needed to understand the chemical and physical relationships between hot corinos and protostellar disks.

In this work, we present ALMA observations of five protostellar disk candidates in the Serpens cluster.  For one source, Ser-emb 1, the chemical and physical structure was studied in detail in \citet{Martin2019}.  Here, we aim to characterize the complex organic chemistry in all five sources in order to further our understanding of the chemical evolution of protostars on small scales.  Section 2 describes our source selection, line targets, and ALMA observations.  Section 3 presents the observed morphologies for each source as well as organic molecule detections.  Additionally, we derive column densities for CH$_3$OH using the population diagram method, and for other species by assuming a rotational temperature.  In Section 4 we discuss the frequency of hot corino detection in our sample of disk candidates.  We also compare the CH$_3$OH-normalized column density ratios of organics in the Serpens sources with measurements in other hot corinos, low-mass protostellar envelopes, and Solar System comets.

\section{Observations \label{sec:obs}}
\subsection{Observational details}
The source sample consists of five low-mass protostars in the Serpens cluster.  Each target source shows non-zero flux at $>$50k$\lambda$ $uv$ distances measured for the 230 GHz continuum, which may be due to the presence of a protostellar disk \citep{Enoch2011}.  Source properties and candidate disk mass estimates are listed in Table \ref{tab:srcdat}.  For all sources we assume a distance of 436 $\pm$ 9 pc \citep{Oritz2018}.

        \begin{table*}\centering
        \footnotesize
        \begin{tabular}{lcccccccc}
        \hline
	\noalign{\vskip 2mm}
        Source & R.A. & Decl. & Class & $T_{bol}$ & $L_{bol}$ & $M_{env}$ & Est. $M_{disk}$ $^a$\\
         & (J2000) & (J2000) && (K) & ($L_\odot$) & ($M_\odot$) & ($M_\odot$) \\
         \noalign{\vskip 0.1mm}
         \hline
         \hline
        \noalign{\vskip 1mm}
	Ser-emb 1 & 18:29:09.1 & 0:31:30.9 & 0 & 39 [2] & 4.1 [0.3] & 3.1 [0.05] & 0.28 \\
	Ser-emb 7 & 18:28:54.1 & 0:29:30.0 & 0 & 58 [13] & 7.9 [0.3] & 4.3 [0.4] & 0.15 \\
	Ser-emb 8 & 18:29:48.1 & 1:16:43.7 & 0 & 58 [16] & 5.4$^b$ [6.2] & 9.4 [0.3] & 0.25 \\
	Ser-emb 15 & 18:29:54.3 & 0:36:00.8 & I & 101 [43] & 0.4 [0.6] & 1.3 [0.1] & 0.15 \\
	Ser-emb 17 & 18:29:06.2 & 0:30:43.1 & I & 117 [21] & 3.8 [3.3] & 3.6  [0.4] & 0.15 \\
        \noalign{\vskip 2mm}
        \hline
        \end{tabular}
        \vspace{0.08in}
        \caption{Source properties, taken from \citet{Enoch2009} and \citet{Enoch2011}. $^a$Uncertainties are at least $\pm$50\%. $^b$Likely an underestimate. \label{tab:srcdat}}
        \end{table*}

The targets were observed during ALMA Cycle 3 from May to June of 2016, as part of project 2015.1.00964.S (PI: K. \"Oberg).  Observations were taken in two different Band 6 spectral setups ranging from 217 -- 233 GHz and 243 -- 262 GHz.  For the lower-frequency setup, each source was observed for $\sim$18 total minutes in two execution blocks, with 42 antennae and baselines spanning 15 -- 560 m.  J1751+0939 was used for bandpass and flux calibration, Titan for flux calibration, and J1830-0619 for phase calibration.  For the higher-frequency setup, each source was observed for $\sim$21 total minutes in three execution blocks, with 41 antennae and baselines spanning 15 -- 640 m.  J1751+0939 was used for bandpass calibration, Titan for flux calibration, and J1830+0619 for phase calibration.  This project makes use of data from three spectral windows centered on 219.57 GHz, 231.49 GHz, and 244.88 GHz.  The 219 GHz window has a bandwidth of 117 MHz and a channel width of 122 kHz ($\sim$0.16 km/s), and the 231 and 244 GHz windows each have a bandwidth of 1.875 GHz and channel width of 488 kHz ($\sim$0.6 km/s).

       \begin{table*}\centering
       \footnotesize
        \begin{tabular}{lcccc}
        \hline
	\noalign{\vskip 2mm}
        Spectral window & Beam dim. & Beam  & Chan. width & Chan. rms  \\
	center frequency & (") & PA ($^{\rm{o}}$) & (km s$^{-1}$) & (mJy beam$^{-1}$)  \\
        \noalign{\vskip 2mm}
\hline 
 \multicolumn{5}{c}{Ser-emb 1} \\ 
 \hline 
219 GHz & 0.63 $\times$ 0.50 & -68.5 & 0.25 & 5.2  \\ 
231 GHz & 0.57 $\times$ 0.45 & -62.5 & 0.63 & 1.9  \\ 
244 GHz & 0.54 $\times$ 0.45 & -59.7 & 0.60 & 2.3  \\ 
\hline 
 \multicolumn{5}{c}{Ser-emb 7} \\ 
 \hline 
219 GHz & 0.61 $\times$ 0.50 & -69.5 & 0.25 & 5.3  \\ 
231 GHz & 0.57 $\times$ 0.45 & -61.8 & 0.63 & 2.0  \\ 
244 GHz & 0.53 $\times$ 0.45 & -59.9 & 0.60 & 2.2  \\ 
\hline 
 \multicolumn{5}{c}{Ser-emb 8} \\ 
 \hline 
219 GHz & 0.63 $\times$ 0.51 & -66.8 & 0.25 & 5.7  \\ 
231 GHz & 0.57 $\times$ 0.45 & -60.9 & 0.63 & 2.0  \\ 
244 GHz & 0.55 $\times$ 0.46 & -63.5 & 0.60 & 2.2  \\ 
\hline 
 \multicolumn{5}{c}{Ser-emb 15} \\ 
 \hline 
219 GHz & 0.67 $\times$ 0.50 & -65.7 & 0.25 & 5.4  \\ 
231 GHz & 0.58 $\times$ 0.45 & -62.2 & 0.63 & 1.9  \\ 
244 GHz & 0.54 $\times$ 0.45 & -59.0 & 0.60 & 2.2  \\ 
\hline 
 \multicolumn{5}{c}{Ser-emb 17} \\ 
 \hline 
219 GHz & 0.63 $\times$ 0.48 & -64.6 & 0.25 & 5.3  \\ 
231 GHz & 0.57 $\times$ 0.45 & -61.7 & 0.63 & 1.9  \\ 
244 GHz & 0.54 $\times$ 0.45 & -58.1 & 0.60 & 2.2  \\ 
        \noalign{\vskip 0.1mm}
        \hline
        \end{tabular}
        \vspace{0.08in}
        \caption{Line observation details \label{tab:obsdat}}
        \end{table*}

\subsection{Analysis}
Initial pipeline calibration of the ALMA data was performed by ALMA/NAASC staff with CASA versions 4.5.3 (lower-frequency setup) and 4.7.0 (higher-frequency setup).  An additional 1--2 rounds of phase self-calibration were performed for each spectral window using the line-free continuum, followed by continuum subtraction.

In addition to the C$^{18}$O 2--1 line in the 219 GHz spectral window, we searched for spectral lines of organic molecules commonly detected in protostars within the 231 GHz and 244 GHz spectral windows.  Spectral line parameters are taken from the JPL \citep{Pickett1998} and CDMS \citep{Muller2001, Muller2005} catalogs.  For CH$_3$OH we searched for lines satisfying A$_{ul}$ $>$10$^{-5}$ s$^{-1}$ and upper energies $<$700 K, and for other COMs we initially searched for lines with A$_{ul}$ $>$10$^{-4}$ s$^{-1}$ and upper energies $<$300 K.  When generating synthetic spectra (Section 3) we include all lines with A$_{ul}$ $>$10$^{-5}$ s$^{-1}$ and upper energies $<$700 K for all COMs.

Image cubes were generated using the \texttt{tclean} task in CASA version 5.4.1, using Briggs weighting with a robust parameter of 0.5.  C$^{18}$O was imaged with a velocity resolution of 0.25 km s$^{-1}$, and all organic lines were imaged at the native spectral resolution.  Clean masks were drawn by hand for detected lines, and for non-detected lines the 10$\sigma$ continuum contour was used.  In addition to hand-masking individual lines, we also cleaned the full data cubes using the automasking task \texttt{auto-multithresh} in \texttt{tclean}.  We used the standard automasking parameters for short baseline 12m line data (sidelobethreshold = 2.0, noisethreshold = 4.5, minbeamfrac = 0.3, negativethreshold = 15.0, lownoisethreshold = 1.5).  We have verified the automasking results by comparing the CH$_3$OH spectra extracted from the auto-masked and hand-masked images, and find they are consistent.  Table \ref{tab:obsdat} shows representative line observation details for each spectral window used in this work.
 
\begin{table*}\centering
\footnotesize
 \begin{tabular}{llccccccc}
 \hline
\noalign{\vskip 2mm}
 Molecule & Transition & Frequency & log(A$_{ul}$) & $g_u$ & $E_u$ & \multicolumn{2}{c}{$Q(T)$} & Refs. \\
 \noalign{\vskip 1mm}
 & &  (GHz) & (s$^{-1}$) & & (K) & 150 K & 250 K &  \\
 \noalign{\vskip 2mm}
 \hline
 \hline
 \noalign{\vskip 1mm}
 C$^{18}$O$^a$ & 2 -- 1 & 219.560 & -6.22 & 5 & 15.8 & 57 & 95 & 1 \\
 \noalign{\vskip 1mm}
 \hline
 \noalign{\vskip 1mm}
 CH$_3$OH$^b$ & 5$_{1,4}$ -- 4$_{1,3}$ A & 243.916 & -4.22 & 44 & 49.7 & 9750 & 26335 &  2 \\ 
 & 10$_{2,8}$ -- 9$_{3,7}$ A & 232.419 & -4.73 & 84 & 165.4 &  &\\ 
 & 10$_{3,7}$ -- 11$_{2,9}$ E & 232.946 & -4.67 & 84 & 190.4 &  &\\ 
 & 9$_{-1,9}$ -- 8$_{-0,8}$ E, $v_t$ = 1& 244.338 & -4.39 & 76 & 395.6 &  &\\ 
 & 18$_{3,16}$ -- 17$_{4,13}$ A & 232.783 & -4.66 & 148 & 446.5 &  &\\
 & 18$_{6,13}$ -- 19$_{5,14}$ \& & 243.397 & -4.70 & 296 & 590.3 &  &\\ 
 & \hspace{0.5em} 18$_{6,12}$ -- 19$_{5,15}$ A $^c$ & & & & & & \\
 & 22$_{3,19}$ -- 22$_{2,20}$ A  & 244.330 & -4.09 & 180 & 636.8 &  &\\ 
 & 23$_{3,20}$ -- 23$_{2,21}$ A & 243.413 & -4.09 & 188 & 690.1 &  &\\ 
 \noalign{\vskip 1mm}
 \hline
 \noalign{\vskip 1mm}
 CH$_3$OCH$_3$$^b$ & 13$_{0,13}$ -- 12$_{1,12}$ AA \& EE $^c$ & 231.988 & -4.04 & 972 & 80.9 & 221973 & 716914 & 3,4 \\ 
  & 23$_{5,18}$ -- 23$_{4,19}$ AA \& EE $^c$ & 243.739 & -4.10 & 1692 & 287.0 &  &\\ 
 \noalign{\vskip 1mm}
 \hline
 \noalign{\vskip 1mm}
 CH$_3$OCHO$^a$ & 19$_{4,15}$ -- 18$_{4,15}$ A &233.227 & -3.74 & 78 & 123.2 & 59073 [1.20]$^d$ & 147269 [1.97]$^d$ & 5--9 \\ 
 & 20$_{4,17}$ -- 19$_{4,16}$ E &244.580 & -3.68 & 82 & 135.0 &  &\\ 
 \noalign{\vskip 1mm}
 \hline
 \noalign{\vskip 1mm}
 NH$_2$CHO$^b$ & 11$_{2,10}$ -- 10$_{2,9}$ & 232.274 & -2.58 & 23 & 78.9 & 3441 & 7461& 10 \\ 
 & 12$_{1,12}$ -- 11$_{1,11}$ & 243.521 & -2.50 & 25 & 79.2 & &\\
 \noalign{\vskip 1mm}
 \hline
 \noalign{\vskip 1mm}
 CH$_2$CO$^b$ & 12$_{1,11}$ -- 11$_{1,10}$ & 244.712 & -3.79 & 75 & 89.4 & 3658 & 7933& 11,12 \\
 \noalign{\vskip 1mm}
 \hline
 \end{tabular}
 \vspace{0.08in}
 \caption{Spectral line data.  $^a$From the JPL catalogue.  $^b$From the CDMS catalogue. $^c$Blended transitions.  $^d$Numbers in brackets represent the vibrational correction factor to the $v_T$ = 0,1 partition function at each temperature. References: [1] \citet{Winnewisser1985} [2] \citet{Xu2008} [3] \citet{Endres2009} [4] \citet{Groner1998} [5] \citet{Ilyushin2009} [6] \citet{Plummer1984} [7] \citet{Oesterling1999} [8] \citet{Maeda2008} [9] \citet{Favre2014} [10] \citet{Kryvda2009} [11] \citet{Brown1990} [12] \citet{Guarnieri2003}
 \label{tab:linedat}}
 \end{table*}

\begin{figure*}
\begin{centering}
	\includegraphics[width=\linewidth]{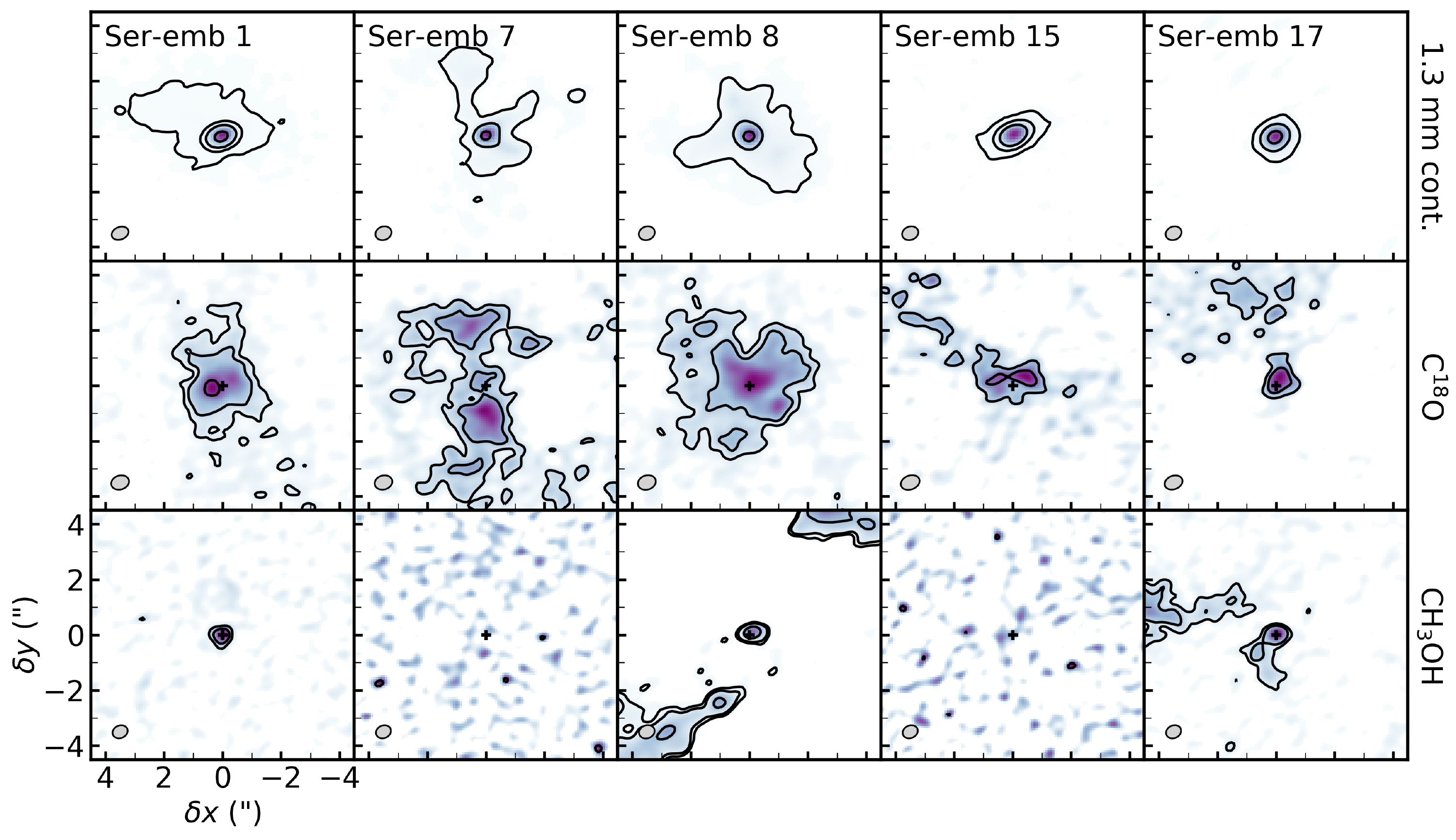}
	\caption{Source overview showing the 1.3 mm dust continuum emission (top), C$^{18}$O 2--1 line emission (middle), and CH$_3$OH 5$_{1,4}$ -- 4$_{1,3}$ line emission (bottom).  Continuum contours are drawn at 5, 30, 100, 400$\times$rms, and line contours are drawn at 5, 10, 30$\times$rms.  Color scales are normalized to each individual image, and emission below a 2$\times$rms threshold is not shown.  The synthesized beam is shown in the bottom left of each panel.  Velocity ranges and rms values for each panel can be found in Appendix \ref{sec:app_mom0}. \label{fig:summary}}
	\end{centering}
\end{figure*}

\section{Results \label{sec:res}}
\subsection{Source overview}
Figure \ref{fig:summary} shows the 1.3 mm dust continuum emission, as well as C$^{18}$O 2--1 and CH$_3$OH 5$_{1,4}$ -- 4$_{1,3}$ line emission (see Table \ref{tab:linedat} for spectral line parameters).  All sources show a compact central dust component, while Ser-emb 1, Ser-emb 7, and Ser-emb 8 additionally have a more extended component.  The C$^{18}$O morphologies show a similar pattern, with more extended gas emission in Ser-emb 1, Ser-emb 7, and Ser-emb 8.  The sources with extended continuum and C$^{18}$O emission are also the least evolved, and it is not surprising that we detect a greater envelope contribution in these sources.  

The CH$_3$OH 5$_{1,4}$ -- 4$_{1,3}$ line ($E_u$ =  50 K) is detected above a 5$\sigma$ level only towards Ser-emb 1, Ser-emb 8, and Ser-emb 17.  Interestingly, the morphology of this transition is different in each source: while all three have a compact central component, Ser-emb 8 also shows a jet-like feature extending out from the central protostar, and Ser-emb 17 shows more amorphous extended emission.  This extended emission is only seen for the 50 K CH$_3$OH line, and all higher-\textit{J} lines show emission only from the compact central component (Figure \ref{fig:com_mom0}).

Our source sample spans a relatively wide range of bolometric temperatures, bolometric luminosities, envelope masses, and estimated disk masses, however there is no clear relationship between these physical properties and the detection of CH$_3$OH in a source (Table \ref{tab:srcdat}).  For each property, the value of one or both sources with CH$_3$OH non-detections falls within the range of values for sources with detections.  We note that there are high uncertainties on the source luminosities for Ser-emb 8 and 17 and on the candidate disk masses for all sources, so better constraints on these properties may reveal some trend with CH$_3$OH emission. There is also no clear relationship between the inferred size of the disk and the presence of CH$_3$OH emission: in the observations of \citet{Enoch2011}, the mm dust emission is compact and unresolved at spatial scales of $\sim$170 AU in Ser-emb 7, Ser-emb 15, and Ser-emb 17, but partially resolved in Ser-emb 1 and Ser-emb 8.  The protostellar neighborhood also does not appear to account for CH$_3$OH emission: none of the sources show evidence for binarity in the \citet{Enoch2011} survey, and Ser-emb 1 is the only source that is isolated more than $\sim$30" from one or two neighboring protostars.  Moreover, Ser-emb 1, 7, and 17 are located near to one another in the Serpens cluster B, while Ser-emb 15 is farther away in cluster B and Ser-emb 8 is in the Main cluster \citep{Enoch2011}.  The presence of CH$_3$OH emission might be influenced by a combination of these and other factors, however, and better characterization of the physical properties of each source as well as their local physical environment is needed.

\subsection{Organic molecule detections}
\begin{figure*}
\begin{centering}
	\includegraphics[width=\linewidth]{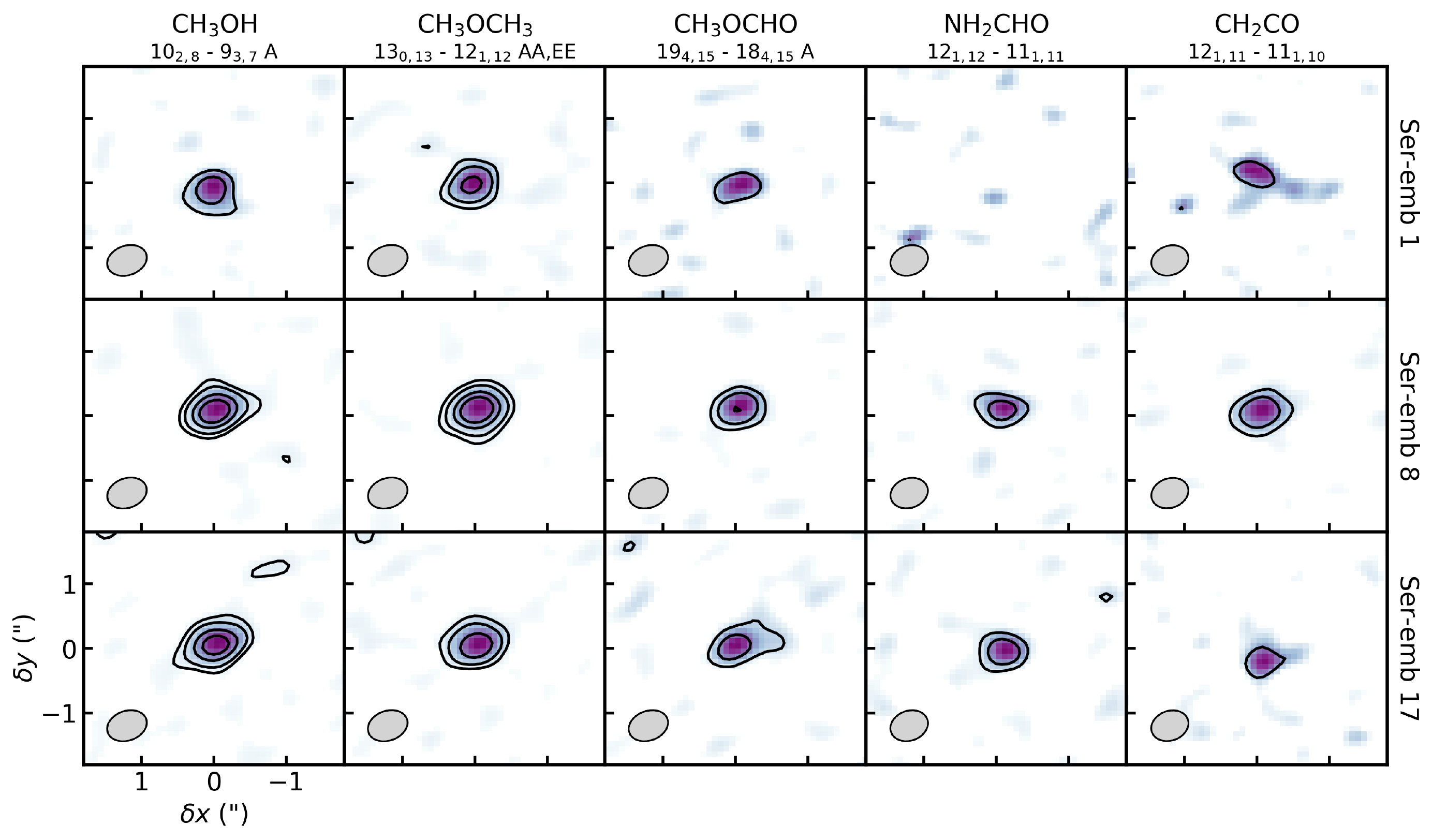}
	\caption{Moment zero maps of organic molecule lines in Ser-emb 1, Ser-emb 8, and Ser-emb 17.  Contours are drawn at 5, 10, 20, 30$\times$rms.  Color scales are normalized to each individual image, and emission below a 2$\times$rms threshold is not shown.  The synthesized beam is shown in the bottom left of each panel.  Velocity ranges and rms values for each panel can be found in Appendix \ref{sec:app_mom0}. \label{fig:com_mom0}}
	\end{centering}
\end{figure*}

The sources with CH$_3$OH detections (Ser-emb 1, Ser-emb 8, and Ser-emb 17) also showed emission from other organic molecules.  We consider a molecule to be detected based on the following criteria.  At least one line must be observed above a 5$\sigma$ level in the moment zero map, and the peak of the extracted spectrum must similarly be $>$5$\sigma$.  For each line, we also check for possible neighbors and do not further consider any lines that are potentially blended with other strong emitters.  Lastly, we search the spectra for other lines of each detected molecule to ensure that any strong transitions are not missing.  Based on these criteria, CH$_3$OCH$_3$, CH$_3$OCHO, and CH$_2$CO were detected towards Ser-emb 1, Ser-emb 8, and Ser-emb 17, while NH$_2$CHO was detected just towards Ser-emb 8 and Ser-emb 17.  We note that the CH$_2$CO detections are based on a single line, however because it is unblended and there are no competing line assignments we are confident in the detection.  Figure \ref{fig:com_mom0} shows moment zero maps for the brightest transition of each of these organic molecules, along with a high-energy ($E_u$ = 165 K) CH$_3$OH line.  In all cases, the emission is unresolved or marginally resolved.  

In several sources we also see evidence for emission from HCOOH, CH$_3$CHO, C$_2$H$_5$OH, and possibly CH$_2$OHCHO.  However, there are too few lines that are both unblended and have sufficient SNR to claim detections.  Higher resolution data and an increased number of line targets per molecule are needed to characterize the emission from these additional species.  

Table \ref{tab:linedat} lists the spectral line information for the transitions that we use for measuring column densities.  Additional CH$_3$OCHO lines are covered in the frequency range of our spectra, however due to line overlap we consider only these non-blended lines for analysis.  We note that the CH$_3$OCHO partition function from JPL includes the $v_T$= 0,1 states, however at hot corino temperatures (200--300 K) other vibrational states will become populated.  We therefore correct the partition function with the vibrational correction factors provided by \citet{Favre2014} via the CMDS catalogue; Table \ref{tab:linedat} lists the non-corrected partition function values and vibrational correction factors at each temperature.  For other molecules, the catalogs already include sufficient vibrationally excited states in the partition function or the vibrational contributions are small at hot corino temperatures; the exception is NH$_2$CHO for which the vibrational contribution is not yet available.

\subsection{Column densities}
\subsubsection{CH$_3$OH \label{sec:cd_ch3oh}}

\begin{figure}
\begin{centering}
	\includegraphics[width=\linewidth]{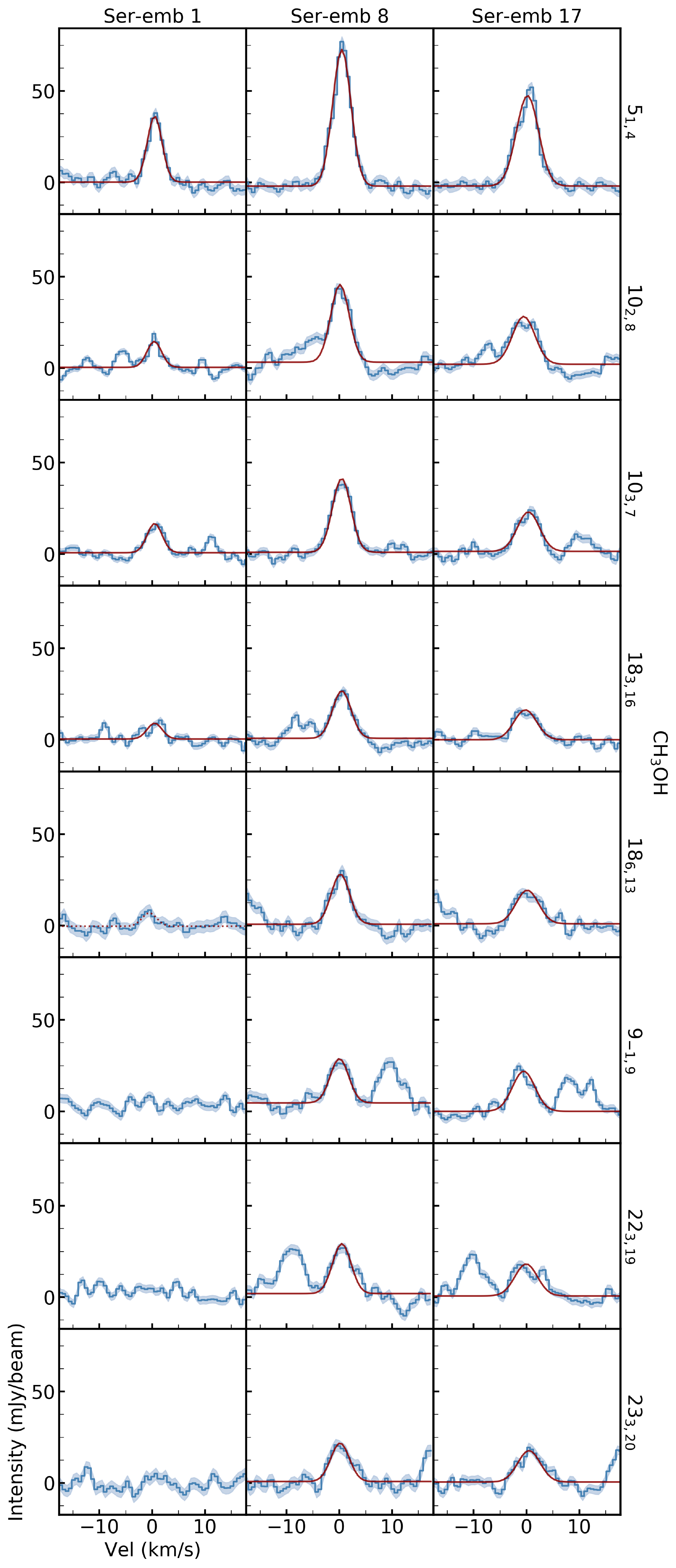}
	\caption{CH$_3$OH spectral lines in sources where CH$_3$OH is detected.  Blue lines show the spectra extracted from the continuum peak pixel, and shaded regions represent the rms.  Red lines show Gaussian fits to the data; a dotted line indicates that the feature is not significant above a 3$\sigma$ level. \label{fig:spec_CH3OH}}
	\end{centering}
\end{figure}

For each source, we extract spectra from a single pixel corresponding to the location of the continuum peak.  Velocity-integrated intensities are measured by fitting a Gaussian to each spectral feature.  For isolated lines we also include an offset term in the fit to allow slight variations in the baseline.  For each source, the line width derived for the CH$_3$OH 5$_{1,4}$ -- 4$_{1,3}$ transition is adopted as a fixed parameter for all other transitions to ensure good fits for weaker or slightly blended lines.  All CH$_3$OH spectral lines and Gaussian fits are shown in Figure \ref{fig:spec_CH3OH}.  Integrated intensities can be found in Table \ref{tab:intfluxes}.  Subsequent uncertainties are propagated based on Gaussian fit uncertainties added in quadrature with a 10\% calibration uncertainty.

Our observations cover 8 CH$_3$OH transitions spanning a range of upper state energies from 50 -- 690 K.  We use the population diagram method to derive column densities and rotational temperatures, taking into account optical depths for each line.  This treatment is adapted from \citet{Goldsmith1999} (see also e.g. \citet{Taquet2015, Loomis2018}) and assumes local thermodynamic equilibrium (LTE) conditions.

The total column density $N_T$ and rotational temperature $T_R$ are related to the upper level population $N_u$ by:

\begin{equation}
\frac{N_u}{g_u} = \frac{N_T}{Q(T_R)} e^{-E_u/T_R}
\label{eq:Nt}
\end{equation}

\noindent Here, $g_u$ is the upper level degeneracy, $Q$ is the molecular partition function, and $E_u$ is the upper state energy in K.  The observed upper level population $N_{u,obs}$ is found from the velocity-integrated surface brightness $\int I_\nu dv$ by:
\begin{equation}
N_{u,obs} = \frac{4\pi\int I_\nu dv}{A_{ul} h c},
\label{eq:Nu}
\end{equation}

\noindent where $A_{ul}$ is the transition Einstein coefficient. If the source does not fill the beam and if the observed lines are optically thick, we find the true upper level population from $N_{u,obs}$ by:

\begin{equation}
N_u = N_{u,obs} C_\tau \frac{\Omega_a}{\Omega_s}  .
\label{eq:Nu_corr}
\end{equation}

\noindent $C_\tau$ is the optical depth correction factor and $\frac{\Omega_a}{\Omega_s}$ is the beam dilution factor, where $\Omega_s$ and $\Omega_a$ are the source and beam solid angles, respectively.  $C_\tau$ is found from:

\begin{equation}
C_\tau = \frac{\tau}{1-e^{-\tau}},
\end{equation}

\noindent and the optical depth $\tau$ is determined by:

\begin{equation}
\tau = \frac{c^3 A_{ul} N_u}{8 \pi \nu^3 \Delta v} (e^{h\nu/kT_R} - 1).
\label{eq:tau0}
\end{equation}

\noindent Here, $c$ is the speed of light, $A_{ul}$ is the Einstein coefficient (lines with a higher $A_{ul}$ are more likely to be optically thick), $\nu$ is the line frequency, $\Delta v$ is the line full width half-maximum, and $k$ is the Boltzmann constant.  For $\Delta v$ we use the FWHM of the CH$_3$OH 5$_{1,4}$--4$_{1,3}$ line in each source (3.3, 4.0, and 4.9 km/s in Ser-emb 1, 8, and 17, respectively).

The degree of beam dilution in our observations is uncertain given that the sources are unresolved or barely resolved.  We therefore solve for column densities using two bounding cases to represent the range of likely values.  The maximum source size, corresponding to the minimum beam dilution, is found using the deconvolved source sizes (or upper limits) derived in CASA for the CH$_3$OH 165 K or 190 K lines (since the 50 K line traces cooler material in addition to the hot corino).  This gives beam dilution factors of 8.7, 15.1, and 9.3 for Ser-emb 1, 8, and 17, respectively.  The minimum source size, corresponding to the maximum beam dilution, is found by using the power-law temperature profile for protostellar envelopes in \citet{Chandler2000} to estimate the radius beyond which the temperature falls below 100 K:

\begin{equation}
T(r) = 60\Bigg{(}\frac{r}{2\times10^{15}\rm{m}}\Bigg{)}^{-q} \Bigg{(}\frac{L_{\rm{bol}}}{10^5 L_\odot}\Bigg{)}^{q/2} \rm{K},
\end{equation}

\noindent where $q$ = 2/(4 + $\beta$).  Assuming the source luminosities in Table 1 and $\beta$ = 1.5, we find maximum beam dilution factors of 27.7, 21.0, and 29.7 for Ser-emb 1, 8, and 17, respectively.

We fit the observed upper level populations (Equation \ref{eq:Nu}) by generating synthetic upper level populations for all detected CH$_3$OH transitions, with $N_T$ and $T_R$ as free parameters.  Combining Equations \ref{eq:Nt} and \ref{eq:Nu_corr}, the synthetic $N_{u,obs}$ are found from:
\begin{equation}
\frac{N_{u,obs}}{g_u} = \frac{N_T}{Q(T_R)}e^{-E_u/T_R}\frac{1}{C_\tau}\frac{\Omega_s}{\Omega_a}.
\label{eq:pd}
\end{equation}
We use the affine-invariant MCMC package \texttt{emcee} \citep{Foreman-Mackey2013} to sample the posterior distributions.  Additional details on the MCMC fitting can be found in Appendix \ref{sec:app_mcmc}. All spectral line parameters used for this analysis are listed in Table \ref{tab:linedat}.

\begin{figure*}
\begin{centering}
	\includegraphics[width=\linewidth]{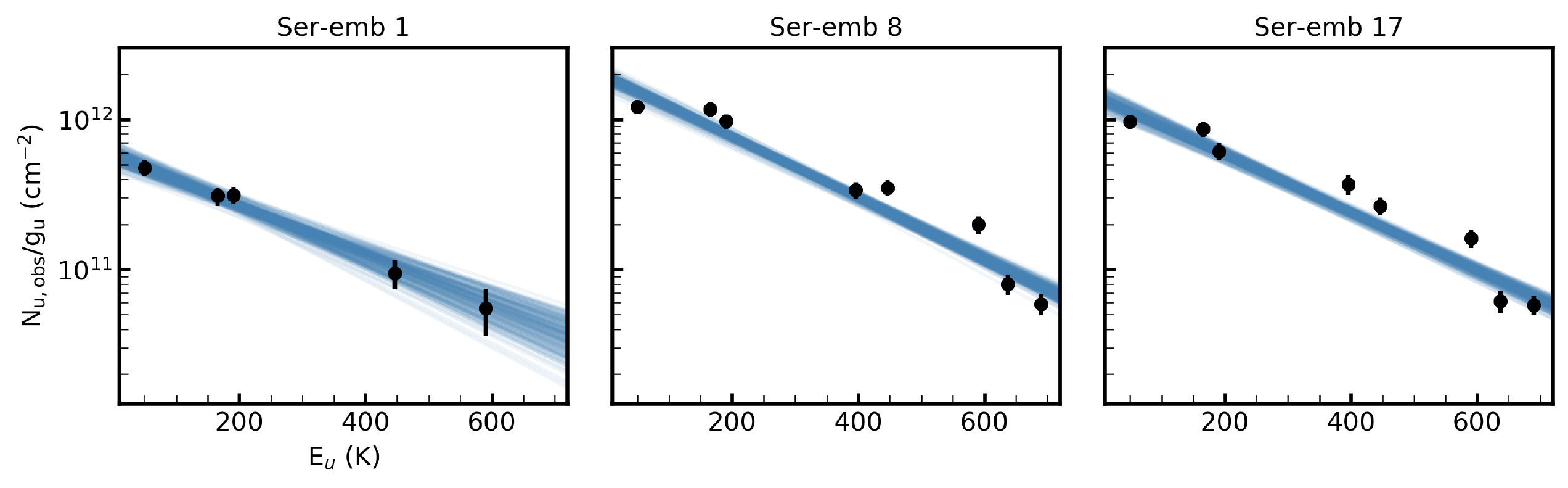}
	\caption{CH$_3$OH population diagrams for Ser-emb 1, Ser-emb 8, and Ser-emb 17.  Data and uncertainties are shown in black, and draws from the fit posteriors are shown in blue. \label{fig:rds}}
	\end{centering}
\end{figure*}

The resulting population diagrams are shown in Figure \ref{fig:rds}.  Table \ref{tab:rd_ch3oh} lists the values derived from this fitting, along with 1$\sigma$ uncertainties.  Across the sample, we derive column densities on the order of 10$^{17}$ cm$^{-2}$ and rotational temperatures of 200--250 K.  Even in the maximum beam dilution cases, in all sources the optical depth of the 50 K line is $<$0.9 and the optical depths of all other lines are $<$0.4.

In hot corino sources where CH$_3$OH lines are optically thick, $^{13}$CH$_3$OH or CH$_3^{18}$OH lines are often used to derive CH$_3$OH column densities \citep{Taquet2015, Jorgensen2016}.  Our observations cover the $^{13}$CH$_3$OH line at 231.818 GHz, however we do not detect this line in any source.  Assuming the maximum beam dilution factor, and a $^{12}$C/$^{13}$C ratio of 70, we obtain 3$\sigma$ upper limits on the CH$_3$OH column density of a few $\times$10$^{19}$ cm$^{-2}$ for our sources.  Thus, these non-detections are consistent with our derived column densities but do not provide any further constraints.

\begin{table*}\centering
\footnotesize
\begin{tabular}{lcccc}
\hline
\noalign{\vskip 2mm}
& \multicolumn{2}{c}{Min. $\frac{\Omega_a}{\Omega_s}$} & \multicolumn{2}{c}{Max. $\frac{\Omega_a}{\Omega_s}$} \\
\noalign{\vskip 1mm}
& N$_T$ (10$^{17}$ cm$^{-2}$) & T$_R$ (K) & N$_T$ (10$^{17}$ cm$^{-2}$) & T$_R$ (K) \\
\noalign{\vskip 2mm}
\hline
\hline
\noalign{\vskip 1mm}
Ser-emb 1 & 1.4 $_{- 0.2}^{+ 0.2}$ & 256 $_{-28}^{+32}$ & 4.5 $_{- 0.7}^{+ 0.6}$ & 256 $_{-28}^{+32}$\\ 
\noalign{\vskip 0.8mm}
Ser-emb 8 & 5.7 $_{- 0.3}^{+ 0.3}$ & 213 $_{-8}^{+8}$ & 8.0 $_{- 0.4}^{+ 0.4}$ & 213 $_{-8}^{+8}$\\ 
\noalign{\vskip 0.8mm}
Ser-emb 17 & 2.8 $_{- 0.1}^{+ 0.1}$ & 224 $_{-9}^{+10}$ & 9.0 $_{- 0.5}^{+ 0.5}$ & 224 $_{-9}^{+9}$\\ 
\noalign{\vskip 1mm}
\hline
\end{tabular}
\caption{CH$_3$OH column densities and rotational temperatures and 1$\sigma$ uncertainties derived from population diagram fitting, assuming minimum and maximum beam dilution factors as described in the text.}\label{tab:rd_ch3oh}
\end{table*}

\subsubsection{Other organics \label{sec:cd_other}}
For all additional organics, detected lines are fitted with Gaussians as described previously for CH$_3$OH.  Again, we adopt the line width of the CH$_3$OH 5$_{1,4}$ -- 4$_{1,3}$ transition for all other molecular lines observed towards a given source.  Spectral line fits for all detected lines can be found in Appendix \ref{sec:app_specfits}, and integrated intensities are listed in Table \ref{tab:intfluxes}.

    \begin{table*}\centering
    \footnotesize
    \begin{tabular}{lccccccc}
    \hline
    \noalign{\vskip 2mm}
     Molecule & Line & \multicolumn{2}{c}{Ser-emb 1} &  \multicolumn{2}{c}{Ser-emb 8}  & \multicolumn{2}{c}{Ser-emb 17} \\
      & & Int. intensity  & Beam dim. & Int. intensity & Beam dim. & Int. intensity & Beam dim.\\
      && (mJy beam$^{-1}$ & (") & (mJy beam$^{-1}$ & (")  & (mJy beam$^{-1}$ & (") \\
      && km s$^{-1}$) & & km s$^{-1}$) & & km s$^{-1}$) & \\
\noalign{\vskip 2mm}
\hline
\hline
\noalign{\vskip 1mm}
CH$_3$OH & 5$_{1,4}$ & 127 $\pm$ 15 & 0.54 $\times$ 0.45 & 325 $\pm$ 34 & 0.55 $\times$ 0.46 & 262 $\pm$ 29 & 0.54 $\times$ 0.45 \\ 
 & 10$_{2,8}$ & 49 $\pm$ 7 & 0.57 $\times$ 0.45 & 187 $\pm$ 21 & 0.57 $\times$ 0.45 & 139 $\pm$ 16 & 0.57 $\times$ 0.45 \\ 
 & 10$_{3,7}$ & 57 $\pm$ 8 & 0.57 $\times$ 0.45 & 178 $\pm$ 19 & 0.57 $\times$ 0.45 & 113 $\pm$ 15 & 0.57 $\times$ 0.45 \\ 
 & 18$_{3,16}$ & 31 $\pm$ 7 & 0.57 $\times$ 0.45 & 114 $\pm$ 14 & 0.57 $\times$ 0.45 & 88 $\pm$ 12 & 0.57 $\times$ 0.45 \\ 
 & 18$_{6,13}$ & 33 $\pm$ 12 & 0.54 $\times$ 0.45 & 120 $\pm$ 16 & 0.54 $\times$ 0.45 & 98 $\pm$ 14 & 0.54 $\times$ 0.45 \\ 
 & 9$_{-1,9}$ & - & -  & 106 $\pm$ 14 & 0.53 $\times$ 0.45 & 117 $\pm$ 18 & 0.54 $\times$ 0.45 \\ 
 & 22$_{3,19}$ & - & -  & 120 $\pm$ 18 & 0.53 $\times$ 0.45 & 93 $\pm$ 16 & 0.54 $\times$ 0.45 \\ 
 & 23$_{3,20}$ & - & -  & 92 $\pm$ 15 & 0.54 $\times$ 0.45 & 92 $\pm$ 13 & 0.54 $\times$ 0.45 \\ 
CH$_3$OCH$_3$ & 13$_{0,13}$ & 102 $\pm$ 12 & 0.57 $\times$ 0.45 & 234 $\pm$ 25 & 0.57 $\times$ 0.45 & 135 $\pm$ 16 & 0.57 $\times$ 0.45 \\ 
 & 23$_{5,18}$ & 74 $\pm$ 10 & 0.54 $\times$ 0.45 & 74 $\pm$ 11 & 0.55 $\times$ 0.46 & 93 $\pm$ 13 & 0.54 $\times$ 0.45 \\ 
CH$_3$OCHO & 19$_{4,15}$ & 35 $\pm$ 10 & 0.57 $\times$ 0.45 & 89 $\pm$ 12 & 0.57 $\times$ 0.45 & 67 $\pm$ 12 & 0.57 $\times$ 0.45 \\ 
 & 20$_{4,17}$ & 48 $\pm$ 9 & 0.53 $\times$ 0.45 & 118 $\pm$ 14 & 0.55 $\times$ 0.46 & 90 $\pm$ 14 & 0.54 $\times$ 0.45 \\ 
NH$_2$CHO & 11$_{2,10}$ & - & -  & 51 $\pm$ 8 & 0.57 $\times$ 0.45 & 44 $\pm$ 11 & 0.57 $\times$ 0.45 \\ 
 & 12$_{1,12}$ & - & -  & 64 $\pm$ 10 & 0.54 $\times$ 0.45 & 68 $\pm$ 11 & 0.54 $\times$ 0.45 \\ 
CH$_2$CO & 12$_{1,11}$ & 30 $\pm$ 7 & 0.53 $\times$ 0.45 & 101 $\pm$ 13 & 0.53 $\times$ 0.45 & 42 $\pm$ 10 & 0.54 $\times$ 0.45 \\ 
    \noalign{\vskip 0.8mm}
    \hline
    \end{tabular}
    \vspace{0.08in}
    \caption{Integrated intensities and uncertainties for spectra extracted from the continuum peak pixel.  For clarity, only upper state quantum numbers are used to identify each line; refer to Table \ref{tab:linedat} for full identifiers. \label{tab:intfluxes}}
    \end{table*}

We estimate column densities by solving equations \ref{eq:Nu} and \ref{eq:pd} assuming optically thin emission and an adopted rotational temperature.  We expect most COMs to share a roughly similar rotational temperature as CH$_3$OH, though some species tend to emit at cooler or warmer temperatures \citep{Jorgensen2016}.  We therefore calculate column densities assuming the CH$_3$OH rotational temperatures derived in the previous section ($T_M$), as well as $T_M$ $\pm$ 75 K.  As for CH$_3$OH, we calculate the range of column density values assuming minimum and maximum beam dilution factors.  The results are listed in Table \ref{tab:rd_proj}. 

\begin{figure*} \centering
\subfloat{
\includegraphics[clip, width=0.95\linewidth]{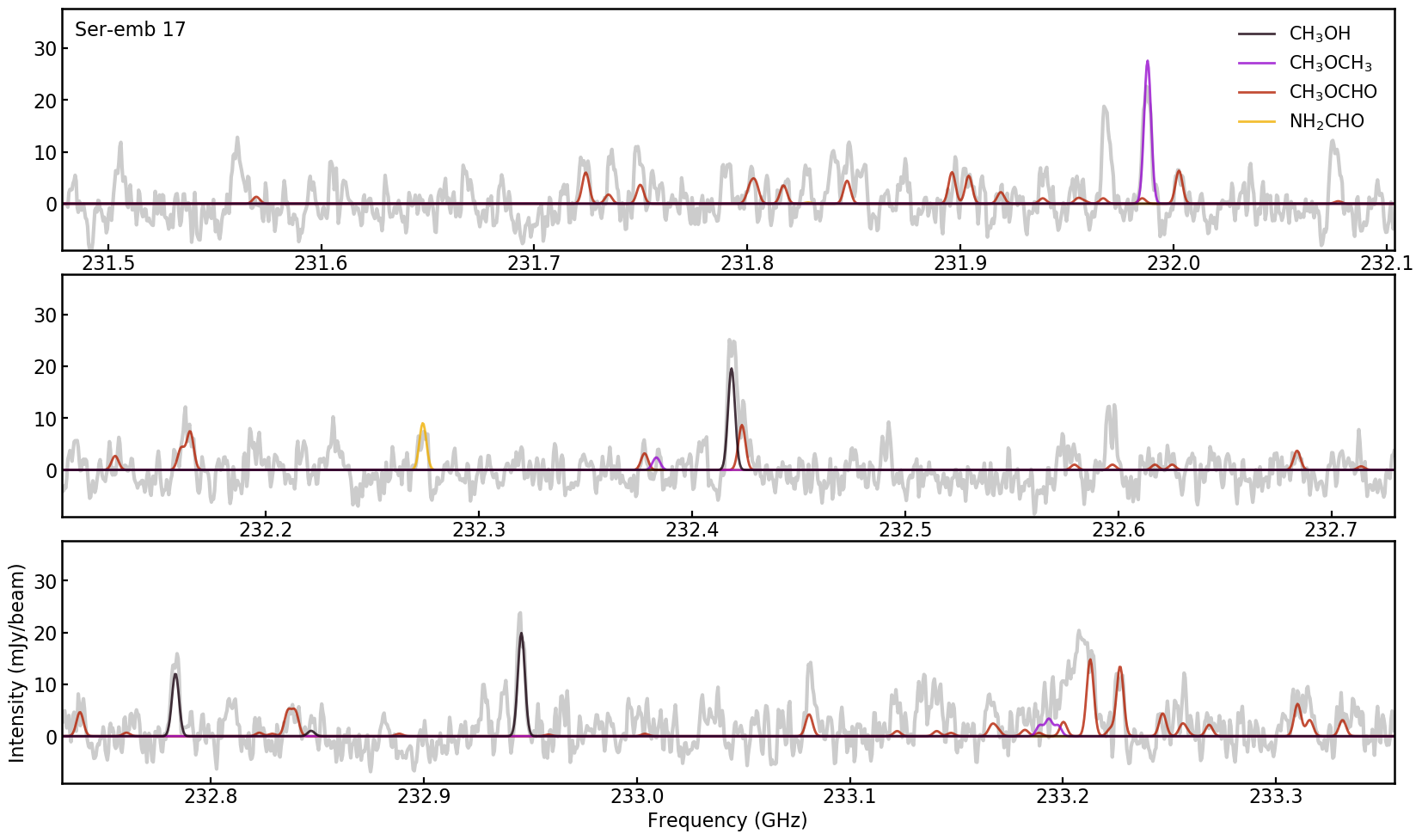}
}
\\
\subfloat{
\includegraphics[clip, width=0.95\linewidth]{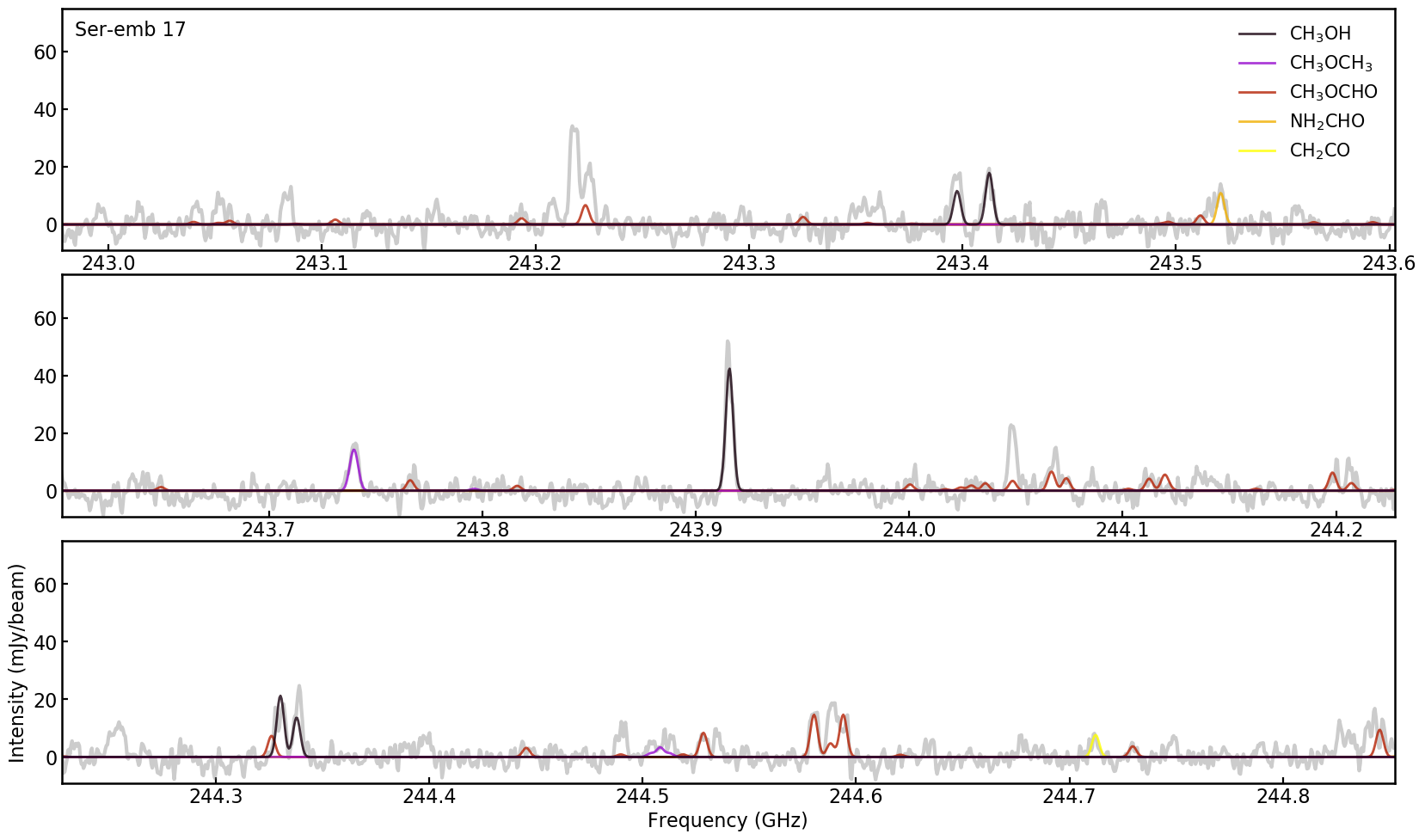}
}
\caption{Full spectrum extracted from the continuum peak pixel in Ser-emb 17 (grey line), along with synthetic spectra of the COMs studied in this work (colored lines) assuming the CH$_3$OH rotational temperature. \label{fig:fullspec_Ser17}}
\end{figure*}

Figure \ref{fig:fullspec_Ser17} shows the complete spectrum extracted from Ser-emb 17 along with synthetic spectra for each organic molecule, calculated with the derived column densities and assuming the CH$_3$OH rotational temperature.  Spectra for all additional sources can be found in Appendix \ref{sec:app_fullspec}.  Synthetic spectra are calculated for all lines with A$_{ul}$ $>$10$^{-5}$ s$^{-1}$ and upper energies $<$700 K.  The optical depths calculated for all COM lines are low ($<$0.3), confirming our assumption of optically thin emission.

\begin{table*}\centering
\footnotesize
\begin{tabular}{lcccccccc}
\hline
\noalign{\vskip 2mm}
&  \multicolumn{2}{c}{CH$_3$OCH$_3$} & \multicolumn{2}{c}{CH$_3$OCHO} & \multicolumn{2}{c}{NH$_2$CHO} & \multicolumn{2}{c}{CH$_2$CO} \\
\noalign{\vskip 1mm}
& \multicolumn{2}{c}{N$_T$ (10$^{17}$ cm$^{-2}$)} & \multicolumn{2}{c}{N$_T$ (10$^{17}$ cm$^{-2}$)} & \multicolumn{2}{c}{N$_T$ (10$^{15}$ cm$^{-2}$)} & \multicolumn{2}{c}{N$_T$ (10$^{15}$ cm$^{-2}$)}\\
\noalign{\vskip 1mm}
& Min. $\frac{\Omega_a}{\Omega_s}$& Max. $\frac{\Omega_a}{\Omega_s}$& Min. $\frac{\Omega_a}{\Omega_s}$& Max. $\frac{\Omega_a}{\Omega_s}$& Min. $\frac{\Omega_a}{\Omega_s}$& Max. $\frac{\Omega_a}{\Omega_s}$& Min. $\frac{\Omega_a}{\Omega_s}$& Max. $\frac{\Omega_a}{\Omega_s}$\\
\noalign{\vskip 2mm}
\hline
\hline
\noalign{\vskip 1mm}
Ser-emb 1 & 1.1 $_{0.7}^{1.2}$  & 3.4 $_{2.2}^{4.0}$  & 1.2 $_{0.5}^{1.7}$  & 3.7 $_{1.7}^{5.3}$ & - & -  & 2.5 $_{1.7}^{2.9}$  & 7.9 $_{5.5}^{9.2}$ \\ 
\noalign{\vskip 1mm}
Ser-emb 8 & 2.2 $_{1.3}^{3.6}$  & 3.0 $_{1.8}^{5.0}$  & 3.1 $_{1.5}^{6.5}$  & 4.3 $_{2.0}^{9.0}$  & 3.1 $_{2.0}^{4.4}$  & 4.3 $_{2.8}^{6.2}$  & 11.8 $_{7.8}^{16.6}$  & 16.4 $_{10.8}^{23.0}$ \\ 
\noalign{\vskip 1mm}
Ser-emb 17 & 1.2 $_{0.8}^{1.8}$  & 3.7 $_{2.4}^{5.7}$  & 1.7 $_{0.8}^{3.5}$  & 5.3 $_{2.4}^{11.1}$  & 1.9 $_{1.2}^{2.7}$  & 6.1 $_{3.9}^{8.6}$  & 3.1 $_{2.1}^{4.3}$  & 9.9 $_{6.5}^{13.8}$ \\ 
\noalign{\vskip 1mm}
\hline
\end{tabular}
\caption{Organic molecule column density estimates.  Column densities are calculated assuming two bounding beam dilution factors, as described in the text.  For each beam dilution factor, column densities are shown for three rotational temperature assumptions: $T_M$ $_{-75 K}^{+ 75 K}$, where $T_M$ is the CH$_3$OH rotational temperature derived in each source.  When multiple lines are detected, the listed column density is the uncertainty-weighted average of the values calculated for each line. \label{tab:rd_proj}}
\end{table*}

\section{Discussion \label{sec:disc}}
\subsection{Frequency of hot corino chemistry around protostellar disk candidates}
In our survey of five protostellar disk candidates, we see evidence for warm, organic-rich, hot-corino-like emission from three of five targeted sources.  This is a high occurrence rate given that only 9 hot corinos have been previously identified in the literature: IRAS 16293-2422 \citep{Cazaux2003}, NGC1333 IRAS 4A \citep{Bottinelli2004}, NGC1333 IRAS 2 \citep{Jorgensen2005}, NGC1333 IRAS 4B \citep{Bottinelli2007}, HH 212 \citep{Codella2016}, B335 \citep{Imai2016}, L483 \citep{Oya2017}, B1b \citep{Lefloch2018}, and SVS 13-A \citep{Lefloch2018}.  Our sources were selected due to their classification as protostellar disk candidates, suggesting that hot corinos and protostellar disks may be evolutionarily or structurally linked.  We emphasize, however, that higher-resolution kinematic studies are needed to verify if these sources indeed host disks.  Interestingly, Ser-emb 17 is just the second Class I source with a hot corino detection \citep{Codella2016, Lefloch2018}; while Class 0 sources represent the majority of hot corino detections, they are clearly not limited to this early evolutionary stage.

\begin{figure*}
\begin{centering}
	\includegraphics[width=\linewidth]{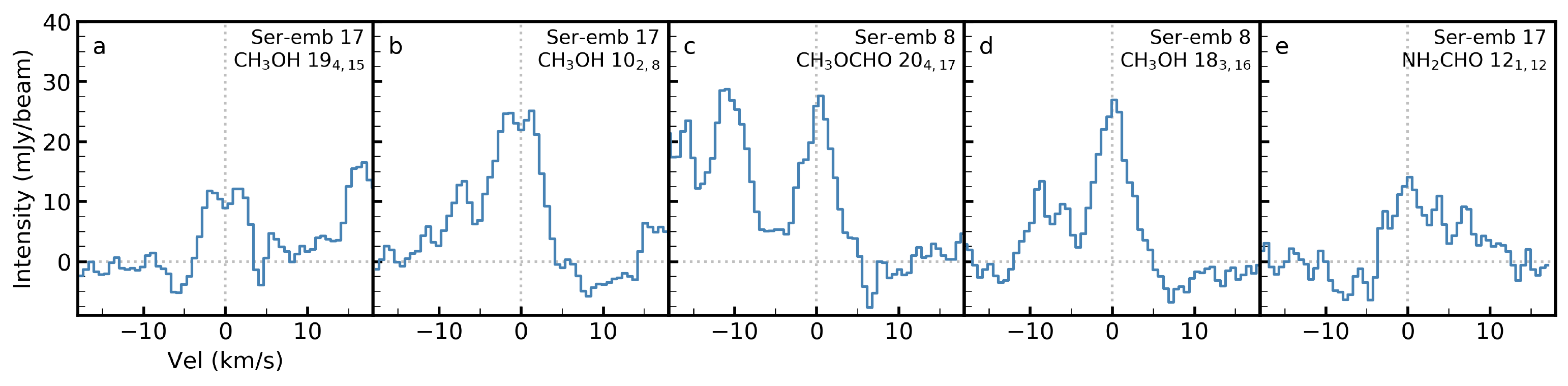}
	\caption{Examples of spectra with line profiles suggestive of rotation (a, b), infall (b, c, d), and outflows (e). For clarity, only upper state quantum numbers are used to identify each line; refer to Table \ref{tab:linedat} for full identifiers. \label{fig:lineprofs}}
	\end{centering}
\end{figure*}

Given the quality of these observations, we cannot at present put strong constraints on whether the organic molecule emission we observe originates from a disk.  However, we do see interesting hints of structure in the line shapes.  Examples of these features are shown in Figure \ref{fig:lineprofs}.  Many organic lines in Ser-emb 17 appear to have a double-peaked line profile, which can be a signature of a rotating disk \citep{Beckwith1993} (though could also be an opacity effect).  Some lines in Ser-emb 8 and Ser-emb 17 also appear to show an inverse P-cygni profile, with an absorption feature occurring at red-shifted wavelengths from the emission peak, indicative of infall \citep{diFrancesco2001, Kristensen2012}.  NH$_2$CHO line profiles in Ser-emb 8 and 17 also show a slight broadening at red-shifted velocities, suggesting the presence of a jet/outflow.  Strong NH$_2$CHO emission in an outflow has been previously reported towards L1157 \citep{Codella2017}.  Clearly, there is evidence for additional structure in these molecular lines that is not resolved in our observations.  Disentangling whether the hot corino, infall, disk, and jet/outflow components are distinct or overlapping in their chemistry and physics requires higher-resolution follow-up observations, and will be key to understanding the connection between protostellar chemistry and disk chemistry.

\subsection{CH$_3$OH column densities \label{sec:disc_tau}}

In other hot corinos that have been studied at high angular resolution (NGC1333 IRAS 2A and 4A, IRAS 16293 B, and L483)\citep{Taquet2015, Jacobsen2018, Jorgensen2018}, derived CH$_3$OH column densities are typically on the order 10$^{19}$ cm$^{-2}$, though a lower value (a few $\times$10$^{17}$ cm$^{-2}$) was found in the incipient hot corino B1b-S \citep{Marcelino2018}.  The CH$_3$OH column densities derived in this work for the Serpens sources are $\sim$10$^{17}$--10$^{18}$cm$^{-2}$, consistent with B1b-S but low compared to the other hot corinos.  We have accounted for beam dilution and optical depth effects as best as possible given the current observations, but additional observations are needed confirm the low CH$_3$OH column densities in the Serpens sources.  Higher-resolution observations that cover additional CH$_3$OH isotopologue lines would enable more robust constraints on the degree to which beam dilution and optical depth effects impact our results. 

If confirmed, the low column densities in the Serpens sources and B1b-S suggest that they are intrinsically weaker/smaller hot corinos, which could be due to their low source luminosities.  This highlights the power of ALMA to detect and characterize new hot corinos that span a broad range of physical properties.

\subsection{Organic abundances across evolutionary stages \label{sec:compare}}
It is of considerable interest to understand how the organic chemistry changes during the evolutionary progression from protostars to protoplanetary disks to planetary systems.  At (sub-)millimeter wavelengths, we can directly probe the gas-phase reservoir of organics, which is also indirectly linked to the ice-phase reservoir through different desorption processes.  In the cold, outer envelopes of low-mass protostars, gas-phase organic molecules likely represent the non-thermal desorption products of grain-surface chemistry.  At the low densities and temperatures of these environments, the formation of saturated organics is not efficient in the gas phase and is thought to proceed exclusively in ice mantles \citep{Horn2004, Geppert2006, Garrod2006}.  Later, in the hot corino stage, gas-phase molecules emit from much warmer and denser regions, where most ice will have sublimated into the gas phase.  Gas-phase compositions therefore reflect thermal ice desorption and possibly additional gas-phase chemistry \citep{Herbst2009, Balucani2015, Skouteris2017, Skouteris2019}.  Finally, we can best explore COM chemistry at the Class II disk stage using cometary measurements, which are made in the gas-phase but are believed to probe pristine ices.  It is not currently clear the extent to which cometary ice compositions are inherited from the interstellar medium, or reprocessed during the assembly of a planetary system.  Protostellar observations suggest that material is altered by heating/shocking during infall and accretion onto the disk \citep{Oya2016}, while various lines of evidence indicate that solid material (both refractory and volatile) in outer Solar system regions is at least partially inherited \citep{Mumma2011, Altwegg2017}.

\begin{figure*}
\begin{centering}
	\includegraphics[width=0.9\linewidth]{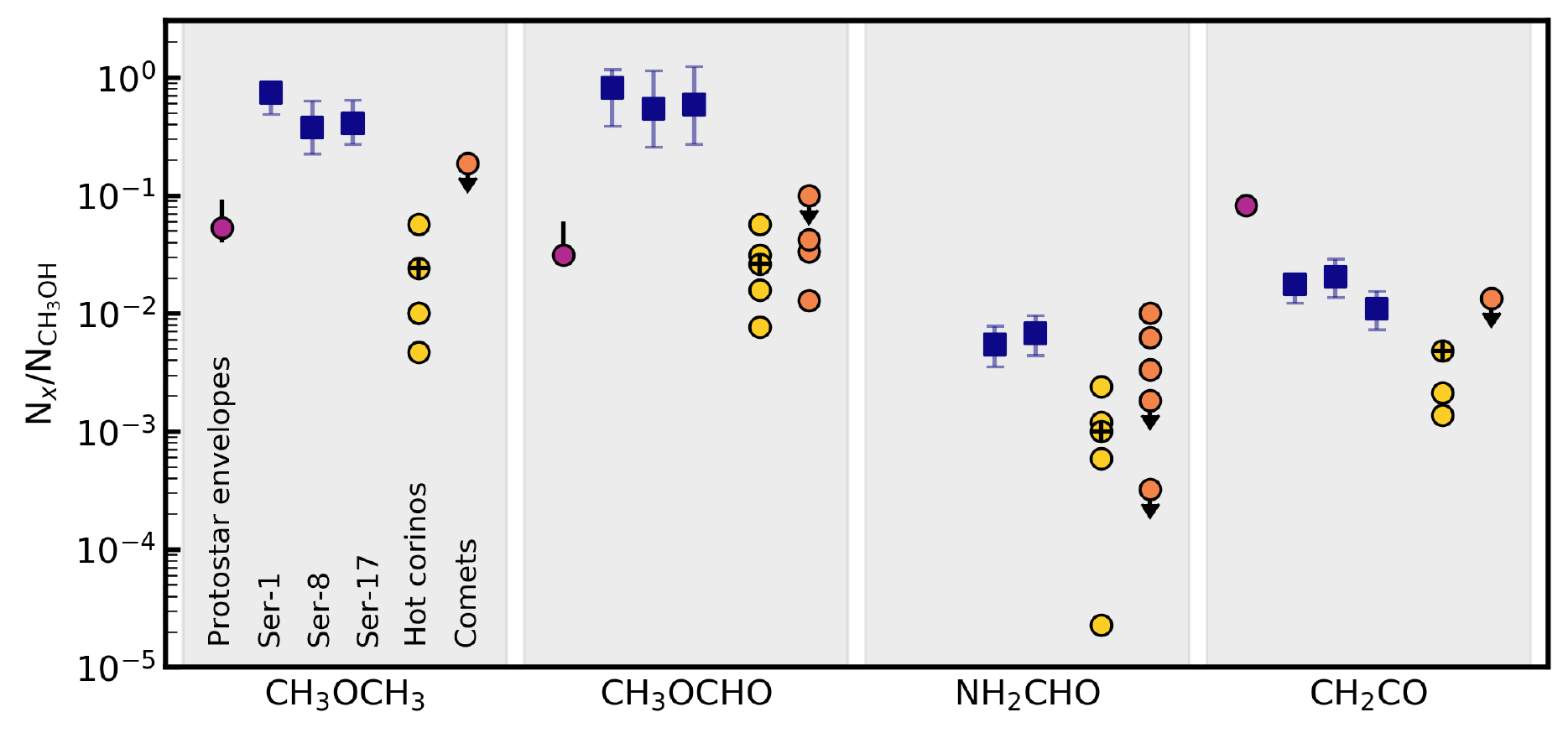}
	\caption{Comparison of organic molecule column density ratios with respect to methanol at different evolutionary stages.  The Ser-emb sources are shown as blue squares.  Column densities are derived for the continuum peak position, assuming the CH$_3$OH rotational temperature $T_M$; error bars show the column densities derived for $T_M$ $\pm$75 K.  Measured abundances in the cold outer envelopes of Class 0/I low-mass protostars are shown as pink dots, hot corinos as yellow dots (+ markers indicate IRAS 16293 B), and Solar System comets as orange dots.  \label{fig:comparison}}
	\end{centering}
\end{figure*}

To explore the organic chemistry in different types of sources, we compare COM/CH$_3$OH column density ratios.  Since CH$_3$OH is predicted to serve as a feedstock molecule for the formation of larger COMs in ice mantles \citep{Oberg2009, Garrod2006, Garrod2008}, the COM/CH$_3$OH ratios can be thought of as a conversion efficiency.  We also note that even in the maximum beam dilution case the optical depths of all lines are $<$1.  This means that uncertainties in the beam dilution factor will cancel out when comparing COM/CH$_3$OH column density ratios.

 Figure \ref{fig:comparison} shows the column density ratios of each organic with respect to CH$_3$OH in the Serpens protostellar disk candidates, along with measurements in a sample of low-mass protostellar envelopes \citep{Bergner2017}; the hot corinos IRAS 16293 B \citep{Jorgensen2018}, L483 \citep{Jacobsen2018}, NGC 1333-IRAS 2A and 4A \citep{Taquet2015}, and Barnard 1b-S \citep{Marcelino2018} \footnote{For the B1b-S hot corino, we have multiplied the $^{13}$CH$_3$OH column densities listed in \citet{Marcelino2018} by 70 to derive a CH$_3$OH column density.}; and the solar system comets Hale-Bopp, Lemmon, Lovejoy, and 67P summer and winter hemispheres \citep{Bockelee2000, Crovisier2004, Biver2014, Biver2015, LeRoy2015}.  For the Serpens sources, organic molecule column densities are calculated assuming the CH$_3$OH rotational temperature $T_M$, and error bars show the values assuming rotational temperatures of $T_M$ $\pm$75 K.

The NH$_2$CHO and CH$_2$CO column density ratios with respect to CH$_3$OH are consistent between the Serpens sources and other hot corinos, while the CH$_3$OCH$_3$ and CH$_3$OCHO ratios in Serpens are enhanced by at least a factor of $\sim$10 compared to the other hot corinos.  There is also at least an order of magnitude spread in the CH$_3$OCH$_3$, CH$_3$OCHO, and NH$_2$CHO ratios across the other hot corinos.  We note that some of this variation may be an artifact of the different angular resolutions and analysis techniques used to derive column densities.  Still, at present it appears that there is a wide range in the abundances of large, oxygen-bearing molecules in different hot corino sources.

Several possible mechanisms could contribute to the observed chemical diversity in hot corinos.  Given that hot corino emission is typically not well resolved, it is possible that different hot corinos host different physical components (e.g. infall, jets, accretion shocks, rotating disks) that alter what molecules are observed in the gas phase.  Alternatively, chemical variation among hot corinos could reflect differences in the amounts of time spent in different physical states.  For instance, the infall rate determines how long ice mantles are heated prior to sublimation, with slower infall allowing more time for the formation of larger organics in lukewarm ices \citep{Garrod2008}.  Time-variable infall chemistry could also cause large differences in the observed organic inventories of different aged sources.  Similarly, if gas-phase chemistry is active following ice sublimation in hot corinos, then the size and age of the hot corino should influence the observed gas-phase abundances.  Interestingly, the Serpens sources are more chemically similar to one another than the other hot corinos, suggesting that the conditions in the local birth cloud may play an important role in setting the protostellar chemistry.

While some or all of these explanations could be at play, we emphasize the need for additional high-resolution and high-sensitivity observations.  Expanding the number of hot corinos for which we have resolved observations of organic molecule emission will be key to understanding the variation in hot corino chemistry.   Additionally, while we have modeled the CH$_3$OH optical depths in the Serpens sources as well as the data will allow, observations of minor CH$_3$OH isotopologues and/or a wider set of lines is needed to confirm the high COM to CH$_3$OH ratios.  

As shown in Figure \ref{fig:comparison}, previously studied hot corinos are chemically similar both to cold protostellar envelopes and to Solar system comets, suggestive of chemical inheritance throughout these evolutionary stages.  However, the Serpens observations show that the hot corino stage is more chemically variable than previously assumed.  This means that care must be taken when using protostellar observations to predict the composition of pre-cometary material in the Solar system.  We need a deeper understanding of what drives chemical diversity at the hot corino stage, and in turn what sources are the best analogs to the young Sun, in order to make inferences about our chemical history.
 
\section{Conclusions}
We have surveyed the organic molecules CH$_3$OH, CH$_3$OCH$_3$, CH$_3$OCHO, NH$_2$CHO, and CH$_2$CO towards five Class 0/I protostellar disk candidates.  Based on our analysis, we conclude the following:
\begin{enumerate}
\item Warm organic molecule emission consistent with hot corino chemistry is detected in three of five targeted sources.  These observations suggest a possible association between hot corinos and protostellar disks.

\item For the three sources with CH$_3$OH detections, we derive column densities of 10$^{17}$--10$^{18}$ cm$^{-2}$ and rotational temperatures of $\sim$200--250 K.  These column densities are on the low end of previously studied hot corinos, highlighting ALMA's capacity to identify smaller or more weakly emitting hot corino sources. 

\item The CH$_3$OH-normalized column density ratios of CH$_3$OCH$_3$ and CH$_3$OCHO in the Serpens sources and other hot corinos span two orders of magnitude.  Local environmental effects and/or time-dependent warm-up chemistry could contribute to this chemical diversity.  

\item High-resolution observations of these and other hot corino sources are required to disentangle whether hot corino emission is associated with a protostellar disk, and to better understand the origins of chemical diversity seen at the hot corino stage.  This will be important for connecting Solar system comet compositions with earlier evolutionary stages.

\end{enumerate}

\section*{Acknowledgements}
This paper makes use of ALMA data, project code 2015.1.00964.S. ALMA is a partnership of ESO (representing its member states), NSF (USA), and NINS (Japan), together with NRC (Canada) and NSC and ASIAA (Taiwan), in cooperation with the Republic of Chile. The Joint ALMA Observatory is operated by ESO, AUI/NRAO, and NAOJ. The National Radio Astronomy Observatory is a facility of the National Science Foundation operated under cooperative agreement by Associated Universities, Inc.

J.B.B acknowledges funding from the National Science Foundation Graduate Research Fellowship under Grant DGE1144152.   This work was supported by an award from the Simons Foundation (SCOL \# 321183, KO).  The group of JKJ is supported by the European Research Council (ERC) under the European Union's Horizon 2020 research and innovation programme through ERC Consolidator Grant ``S4F'' (grant agreement No~646908). Research at Centre for Star and Planet Formation is funded by the Danish National Research Foundation.

This research made use of \texttt{emcee} \citep{Foreman-Mackey2013}, \texttt{NumPy} \citep{VanDerWalt2011}, \texttt{Matplotlib}\citep{Hunter2007}, \texttt{Astropy} \citep{Astropy2013}, and \texttt{SciPy} \citep{Jones2011}.

\newpage

\FloatBarrier

\appendix

\section{Moment zero map parameters}
\label{sec:app_mom0}
\noindent Table \ref{tab:mom0_dat} lists the rms values and velocity ranges used to make the moment zero maps shown in Figures \ref{fig:summary} and \ref{fig:com_mom0}.

\begin{table*}\centering
\footnotesize
\begin{tabular}{lcccccccccc}
\hline
\noalign{\vskip 2mm}
& \multicolumn{2}{c}{Ser-emb 1} & \multicolumn{2}{c}{Ser-emb 7} & \multicolumn{2}{c}{Ser-emb 8} & \multicolumn{2}{c}{Ser-emb 15} & \multicolumn{2}{c}{Ser-emb 17} \\
\noalign{\vskip 1mm}
& Vel. & rms & Vel. & rms & Vel. & rms & Vel. & rms & Vel. & rms   \\
& (km/s) & (mJy) & (km/s) & (mJy) & (km/s) & (mJy) & (km/s) & (mJy) & (km/s) & (mJy) \\
\noalign{\vskip 2mm}
\hline
\hline
\noalign{\vskip 1mm}
C$^{18}$O 2& 6.7--10.5 & 4.1& 6.0--12.0 & 5.4& 6.0--10.7 & 5.6& 8.2--12.5 & 4.4& 4.7--10.5 & 5.7\\ 
CH$_3$OH 5$_{1,4}$& 2.6--11.6 & 6.8& 6.2--11.6 & 4.1& 2.6--14.0 & 6.5& 8.6--12.2 & 3.2& 2.0--13.4 & 7.5\\ 
CH$_3$OH 10$_{2,8}$& 7.0--10.2 & 2.9& ... & ...& 4.5--12.7 & 5.4& ... & ...& 5.1--10.8 & 3.5\\ 
CH$_3$OCH$_3$ 13$_{0,13}$& 6.4--12.1 & 3.7& ... & ...& 3.2--12.1 & 5.5& ... & ...& 4.5--12.1 & 4.3\\ 
CH$_3$OCHO 19$_{4,15}$& 5.7--9.5 & 3.4& ... & ...& 5.7--12.6 & 5.0& ... & ...& 4.4--10.1 & 4.1\\ 
NH$_2$CHO 12$_{1,12}$& 6.8--9.8 & 3.2& ... & ...& 6.2--12.8 & 5.2& ... & ...& 5.0--12.8 & 5.3\\ 
CH$_2$CO 12$_{1,11}$& 5.6--12.1 & 4.7& ... & ...& 4.4--12.1 & 5.4& ... & ...& 4.4--12.1 & 5.1\\ 
\noalign{\vskip 1mm}
\hline
\end{tabular}
\caption{Velocity ranges and rms values used for the integrated intensity maps (Figures \ref{fig:summary} and \ref{fig:com_mom0}) in the main text.  For clarity, only upper state quantum numbers are used to identify each line; refer to Table \ref{tab:linedat} for full identifiers.}\label{tab:mom0_dat}
\end{table*}

\FloatBarrier
\section{Population diagram diagram fitting}
\label{sec:app_mcmc}

\begin{figure*} \centering
\subfloat{
\includegraphics[clip, width=0.45\linewidth]{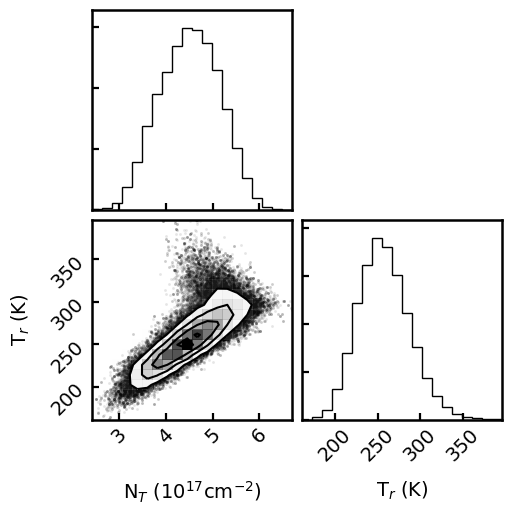}
}
\subfloat{
\includegraphics[clip, width=0.5\linewidth]{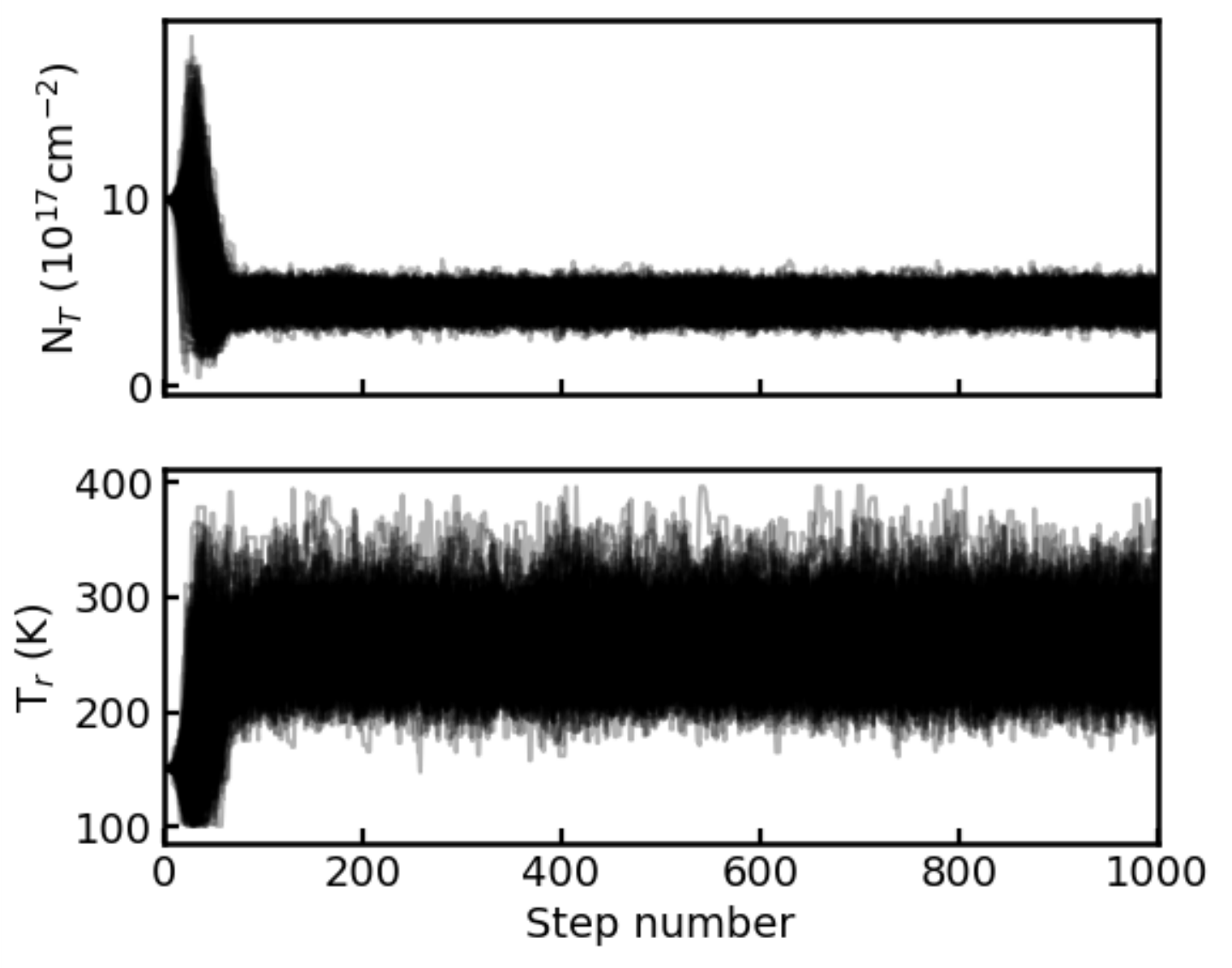}
}
\caption{Ser-emb 1 rotational diagram MCMC fit results.  The corner plot is shown on the left and the walker chain on the right. \label{fig:mcmc_Ser1}}
\end{figure*}

\begin{figure*} \centering
\subfloat{
\includegraphics[clip, width=0.45\linewidth]{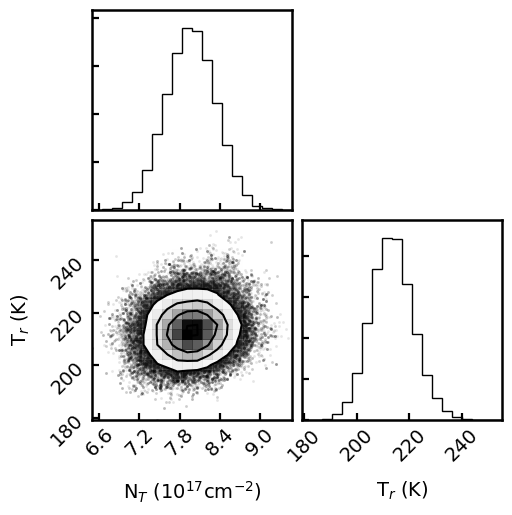}
}
\subfloat{
\includegraphics[clip, width=0.5\linewidth]{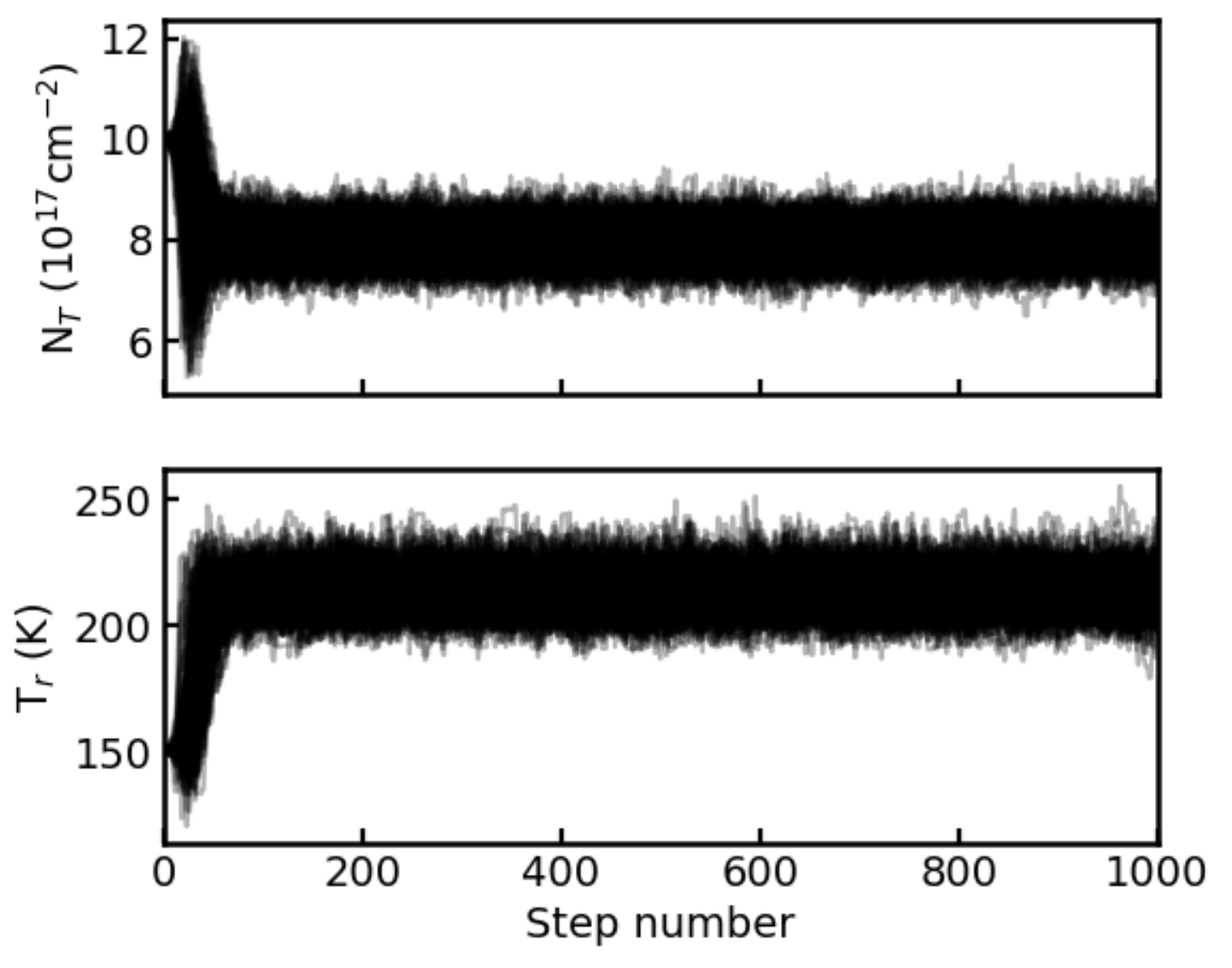}
}
\caption{Ser-emb 8 rotational diagram MCMC fit results.  The corner plot is shown on the left and the walker chain on the right. \label{fig:mcmc_Ser8}}
\end{figure*}

\begin{figure*} \centering
\subfloat{
\includegraphics[clip, width=0.45\linewidth]{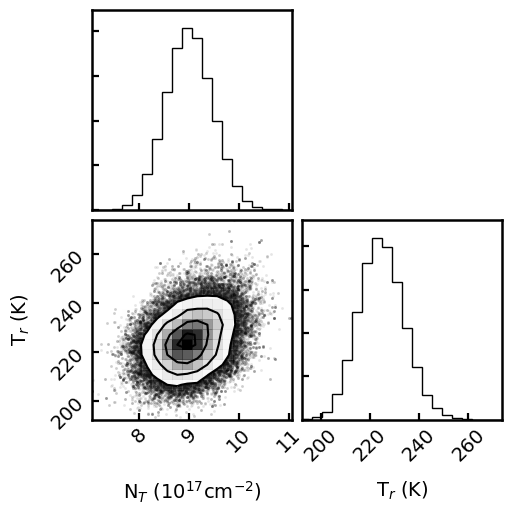}
}
\subfloat{
\includegraphics[clip, width=0.5\linewidth]{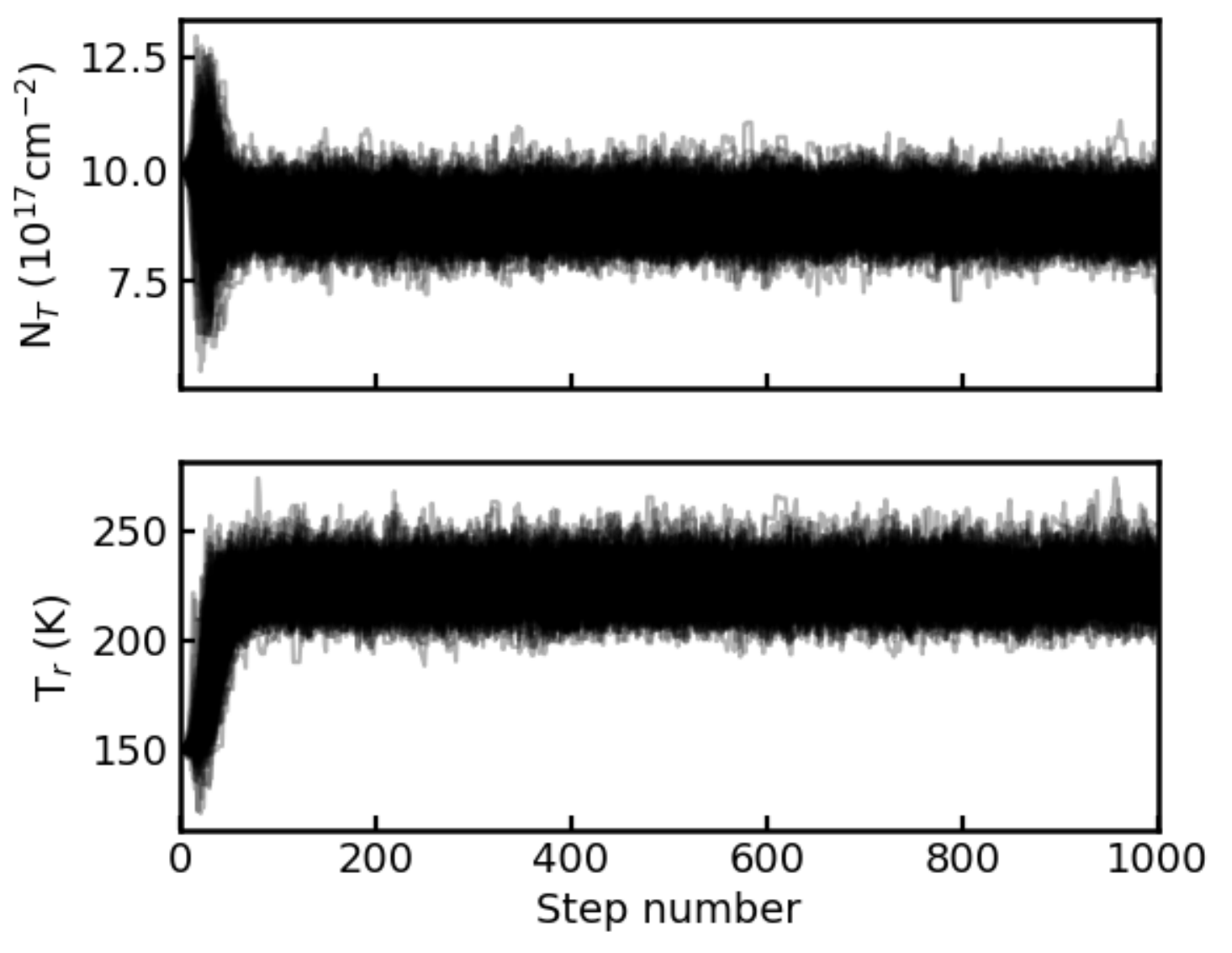}
}
\caption{Ser-emb 17 rotational diagram MCMC fit results.  The corner plot is shown on the left and the walker chain on the right. \label{fig:mcmc_Ser17}}
\end{figure*}

\noindent For the MCMC population diagram fits to CH$_3$OH, we use a flat prior 10$^5$ $<$ N$_T$ $<$ 10$^{20}$ cm$^{-2}$ and 100 $<$ T$_r$ $<$ 400 K.  200 walkers are propagated for 1000 steps, and the samples are well converged.  Walker chains and corner plots for each source are shown in Figures \ref{fig:mcmc_Ser1}--\ref{fig:mcmc_Ser17} for the maximum beam dilution case.

\FloatBarrier
\section{Spectral line fits}
\label{sec:app_specfits}

\begin{figure*}
\begin{centering}
	\includegraphics[width=0.6\linewidth]{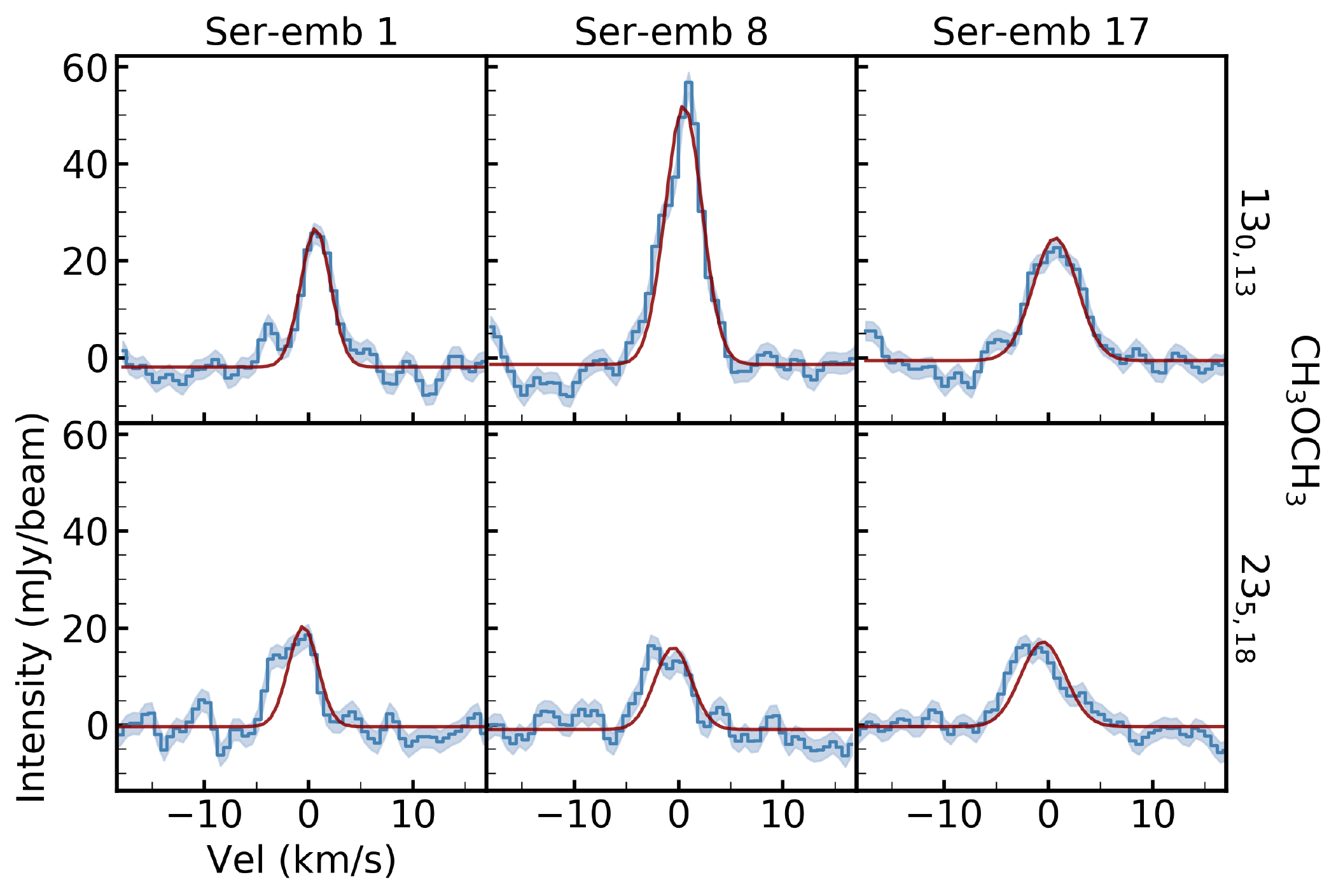}
	\caption{CH$_3$OCH$_3$ spectral lines.  Blue lines show the spectra extracted from the continuum peak pixel, and shaded regions represent the rms.  Red lines show Gaussian fits to the data. \label{fig:spec_CH3OCH3}}
	\end{centering}
\end{figure*}

\begin{figure*}
\begin{centering}
	\includegraphics[width=0.6\linewidth]{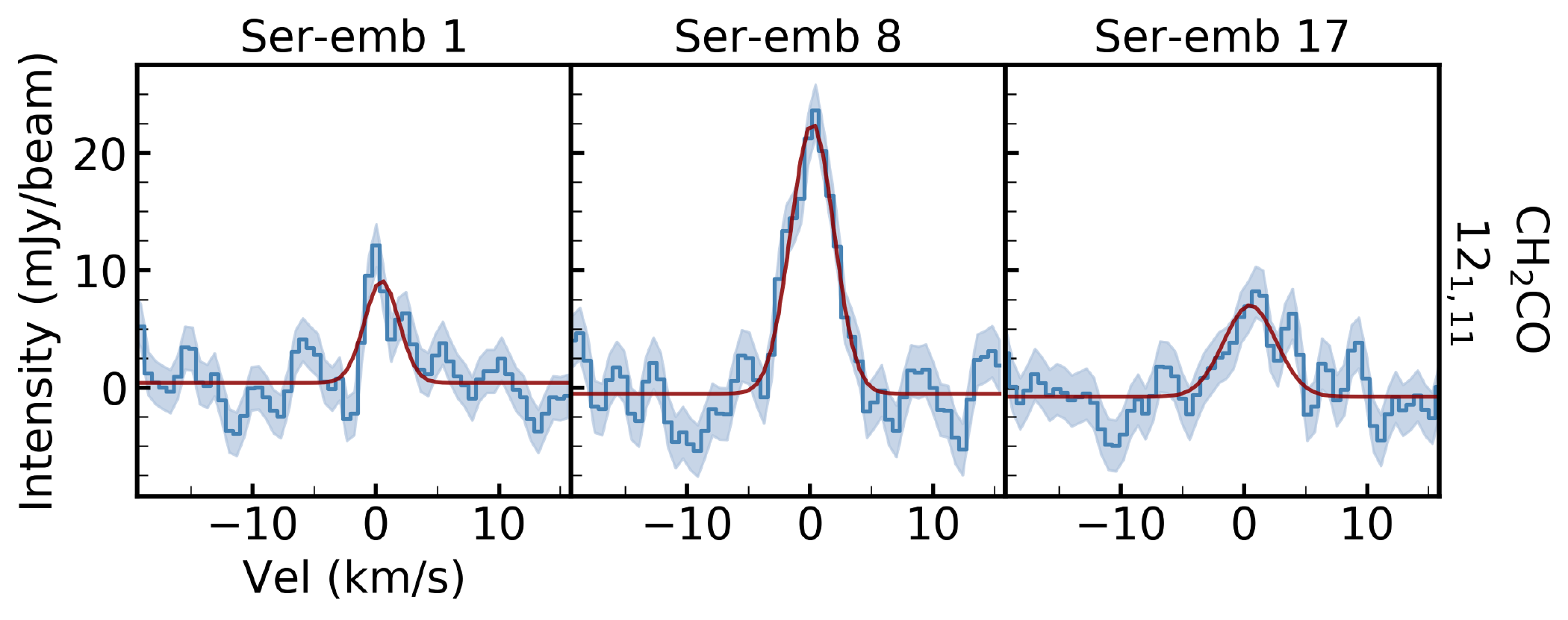}
	\caption{CH$_2$CO spectra lines.  Blue lines show the spectra extracted from the continuum peak pixel, and shaded regions represent the rms.  Red lines show Gaussian fits to the data. \label{fig:spec_CH2CO}}
	\end{centering}
\end{figure*}

\noindent Figures \ref{fig:spec_CH3OCH3}--\ref{fig:spec_CH2CO} show Gaussian fits to the observed lines of each COM, analogous to Figure \ref{fig:spec_CH3OH}.

\begin{figure*}
\begin{centering}
	\includegraphics[width=0.6\linewidth]{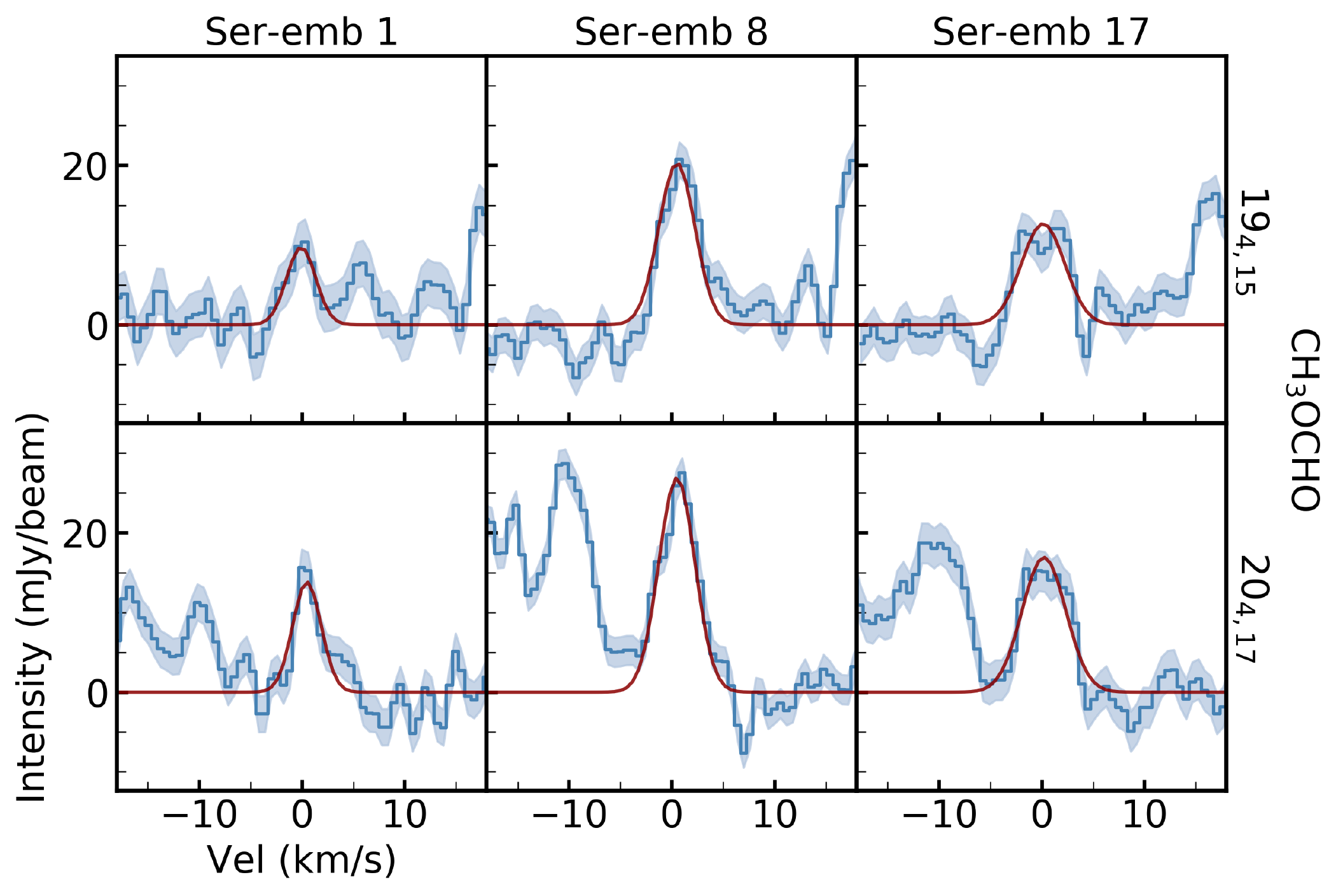}
	\caption{CH$_3$OCHO spectral lines.  Blue lines show the spectra extracted from the continuum peak pixel, and shaded regions represent the rms.  Red lines show Gaussian fits to the data.}
	\end{centering}
\end{figure*}

\begin{figure*}
\begin{centering}
	\includegraphics[width=0.4\linewidth]{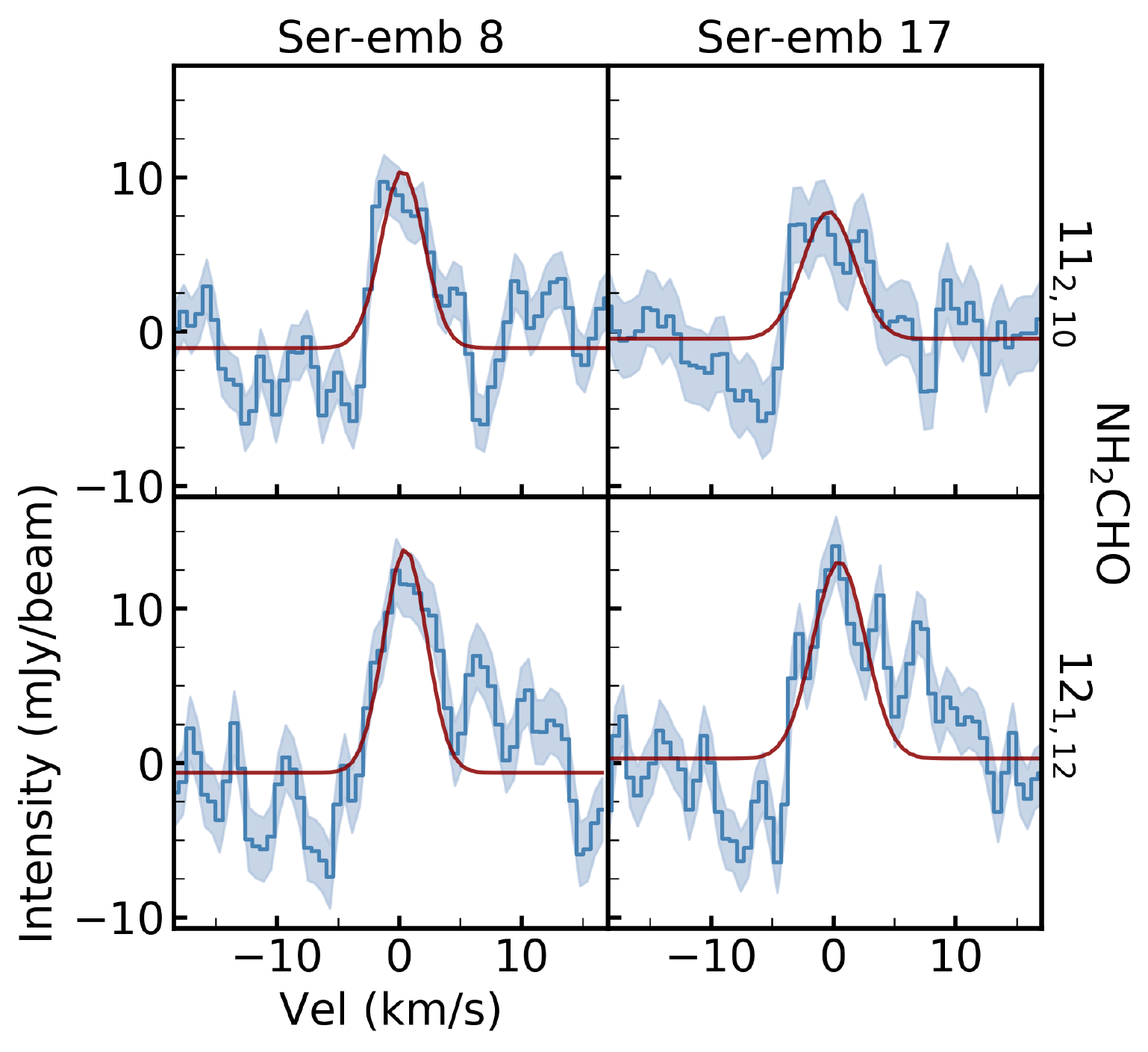}
	\caption{NH$_2$CHO spectral lines.  Blue lines show the spectra extracted from the continuum peak pixel, and shaded regions represent the rms.  Red lines show Gaussian fits to the data.}
	\end{centering}
\end{figure*}

\FloatBarrier
\section{Full spectra}
\label{sec:app_fullspec}

\begin{figure*} \centering
\subfloat{
\includegraphics[clip, width=0.95\linewidth]{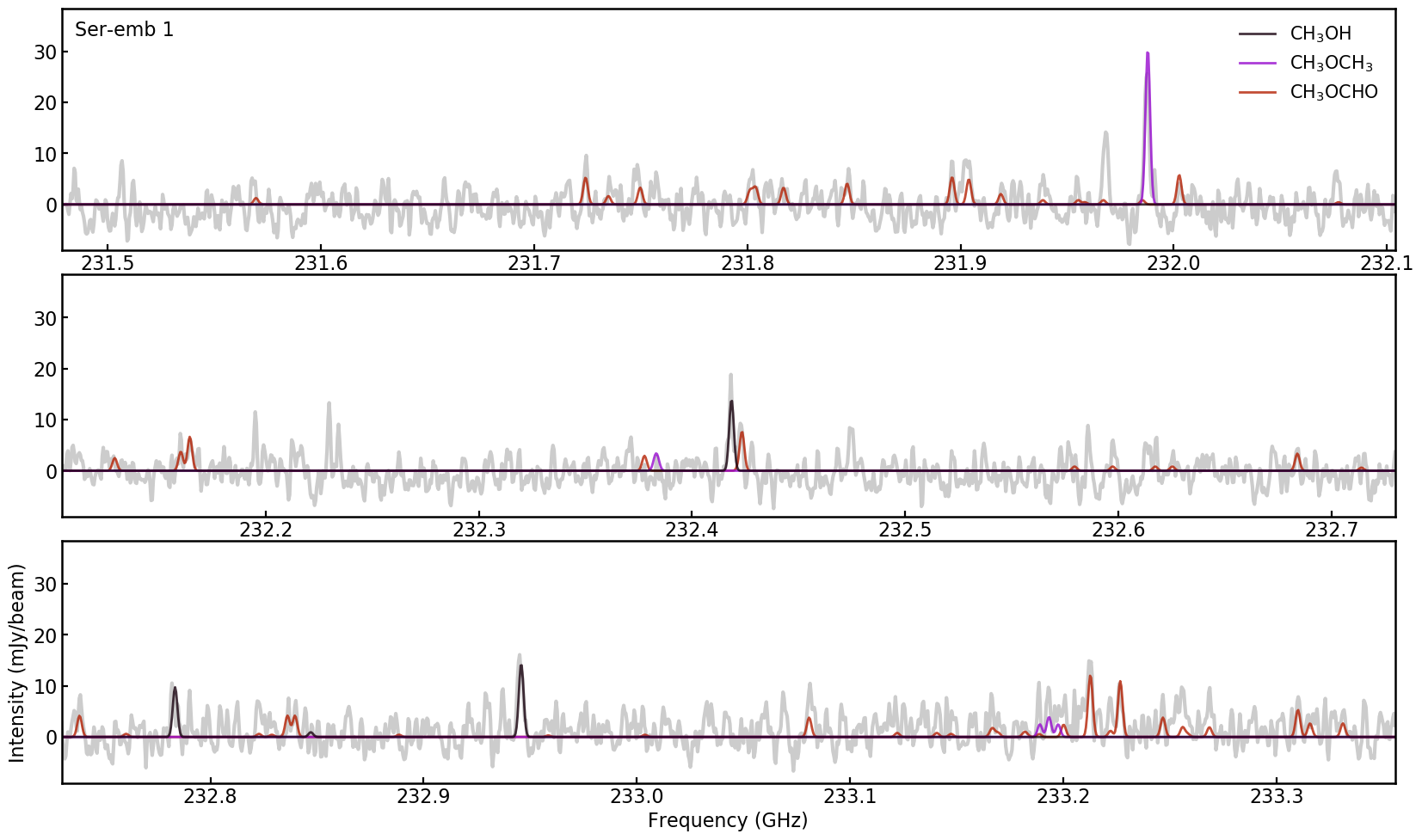}
}
\\
\subfloat{
\includegraphics[clip, width=0.95\linewidth]{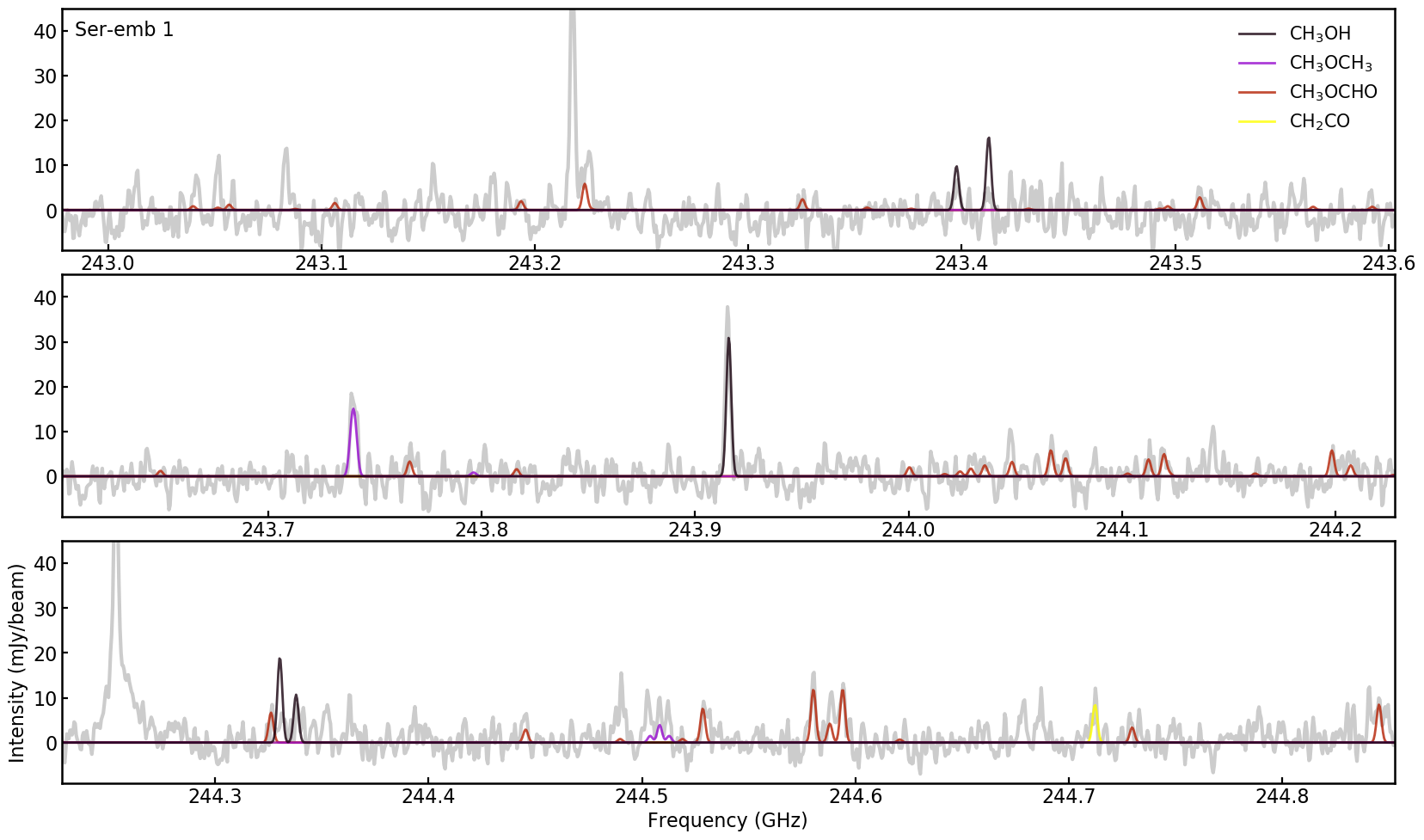}
}
\caption{Full spectrum extracted from the continuum peak pixel in Ser-emb 1 (grey line), along with synthetic spectra of the detected COMs (colored lines).  Spectra are calculated assuming the CH$_3$OH rotational temperature. \label{fig:fullspec_Ser1}}
\end{figure*}

\begin{figure*} \centering
\subfloat{
\includegraphics[clip, width=0.95\linewidth]{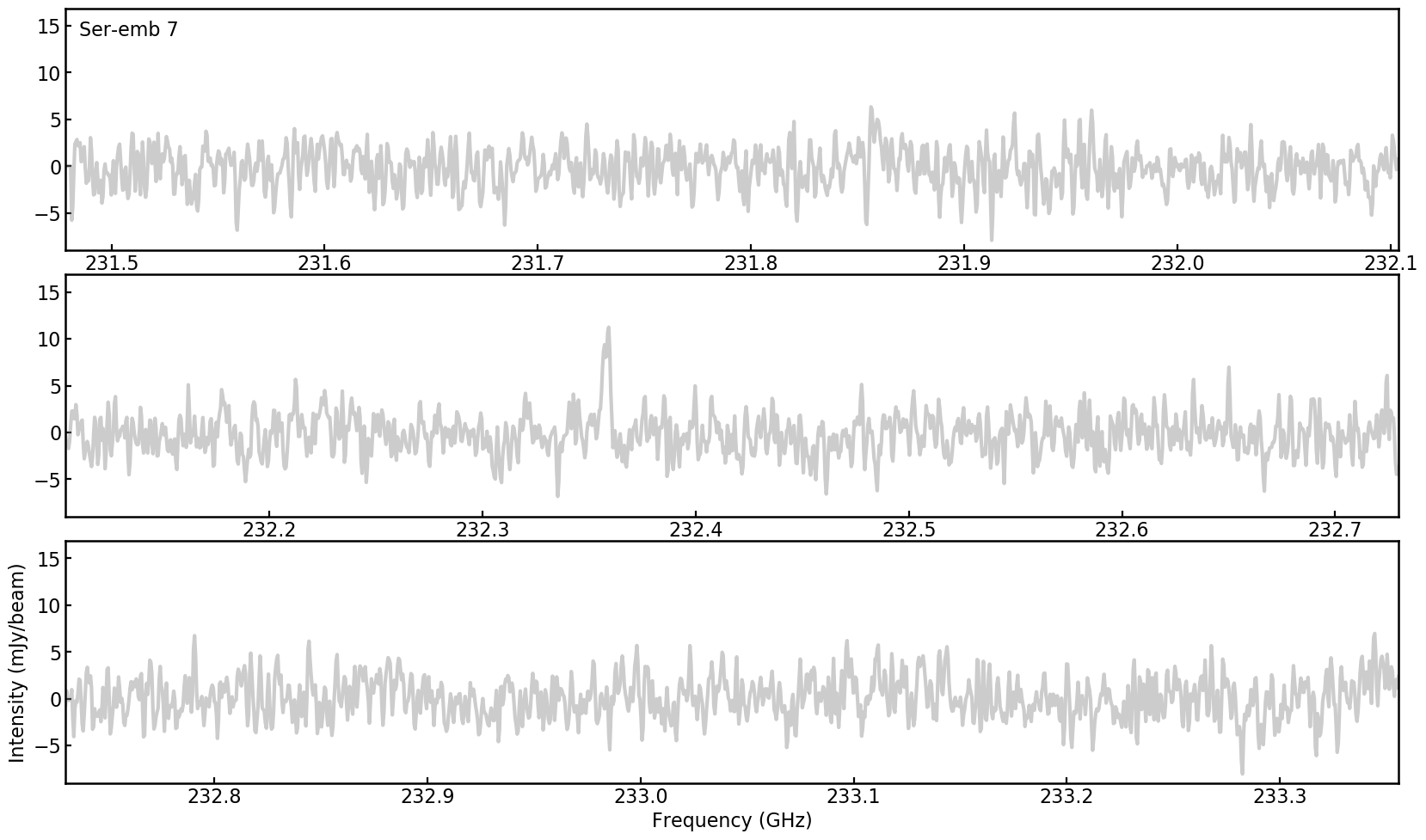}
}
\\
\subfloat{
\includegraphics[clip, width=0.955\linewidth]{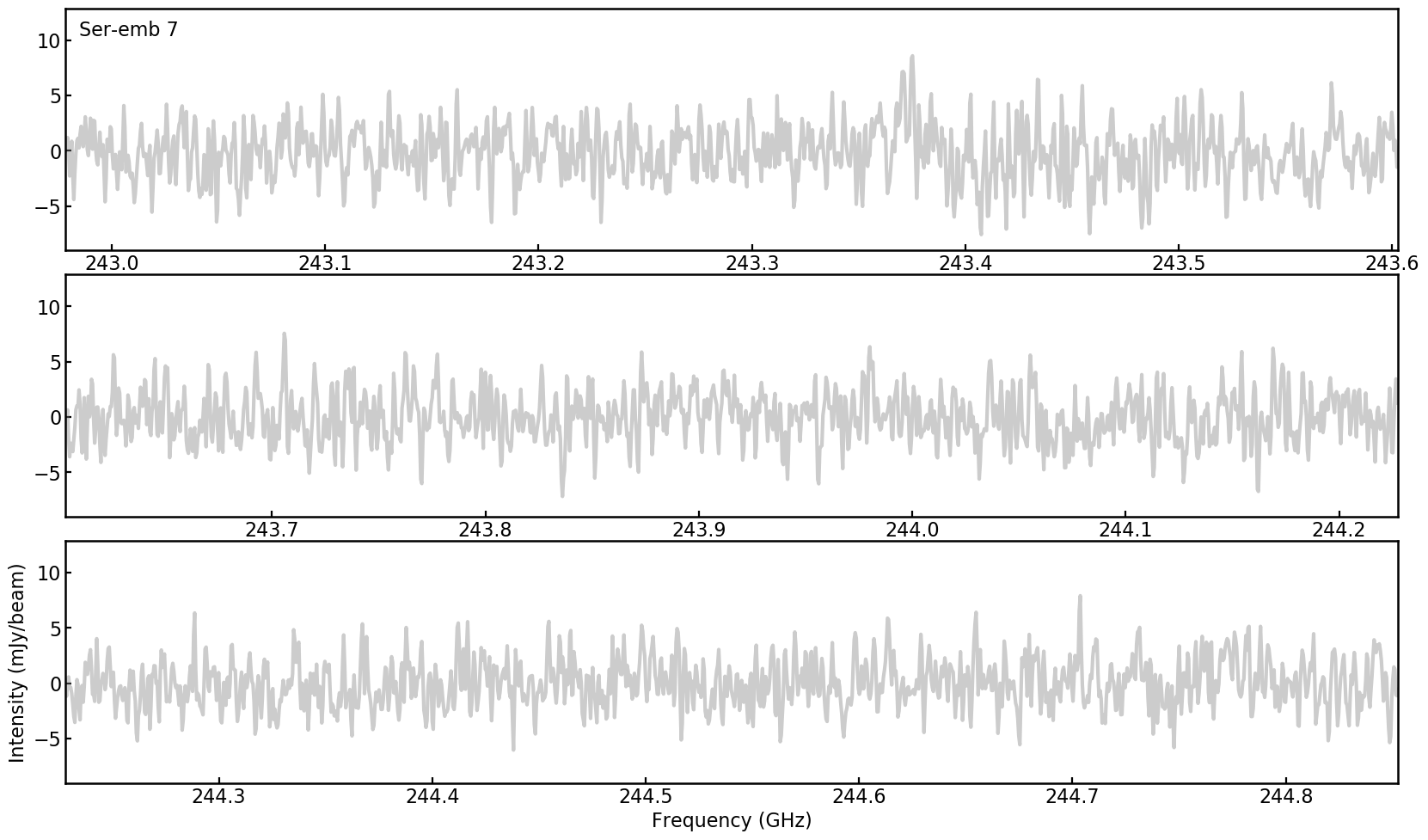}
}
\caption{Full spectrum extracted from the continuum peak pixel in Ser-emb 7 (grey line). \label{fig:fullspec_Ser7} }
\end{figure*}

\begin{figure*} \centering
\subfloat{
\includegraphics[clip, width=0.95\linewidth]{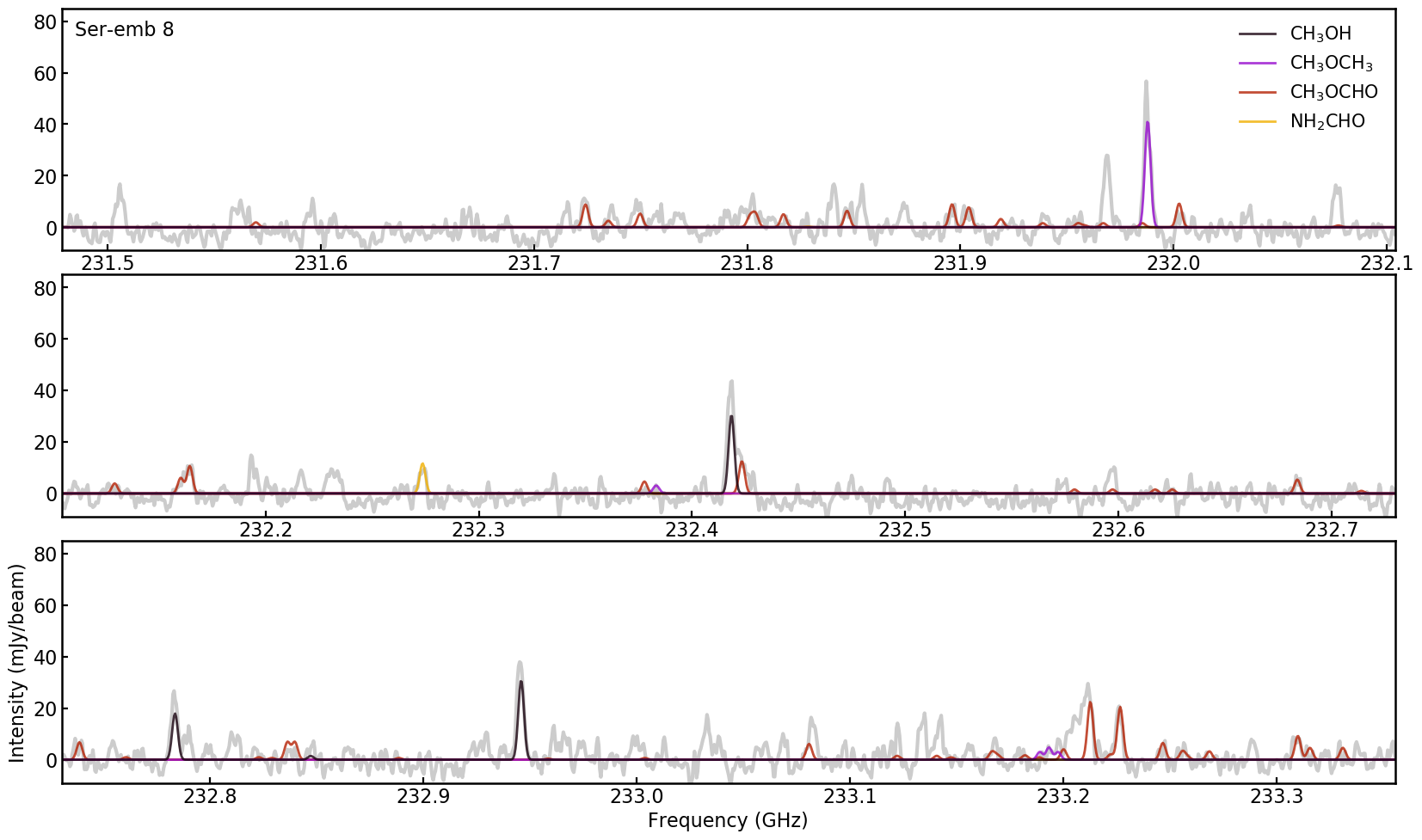}
}
\\
\subfloat{
\includegraphics[clip, width=0.95\linewidth]{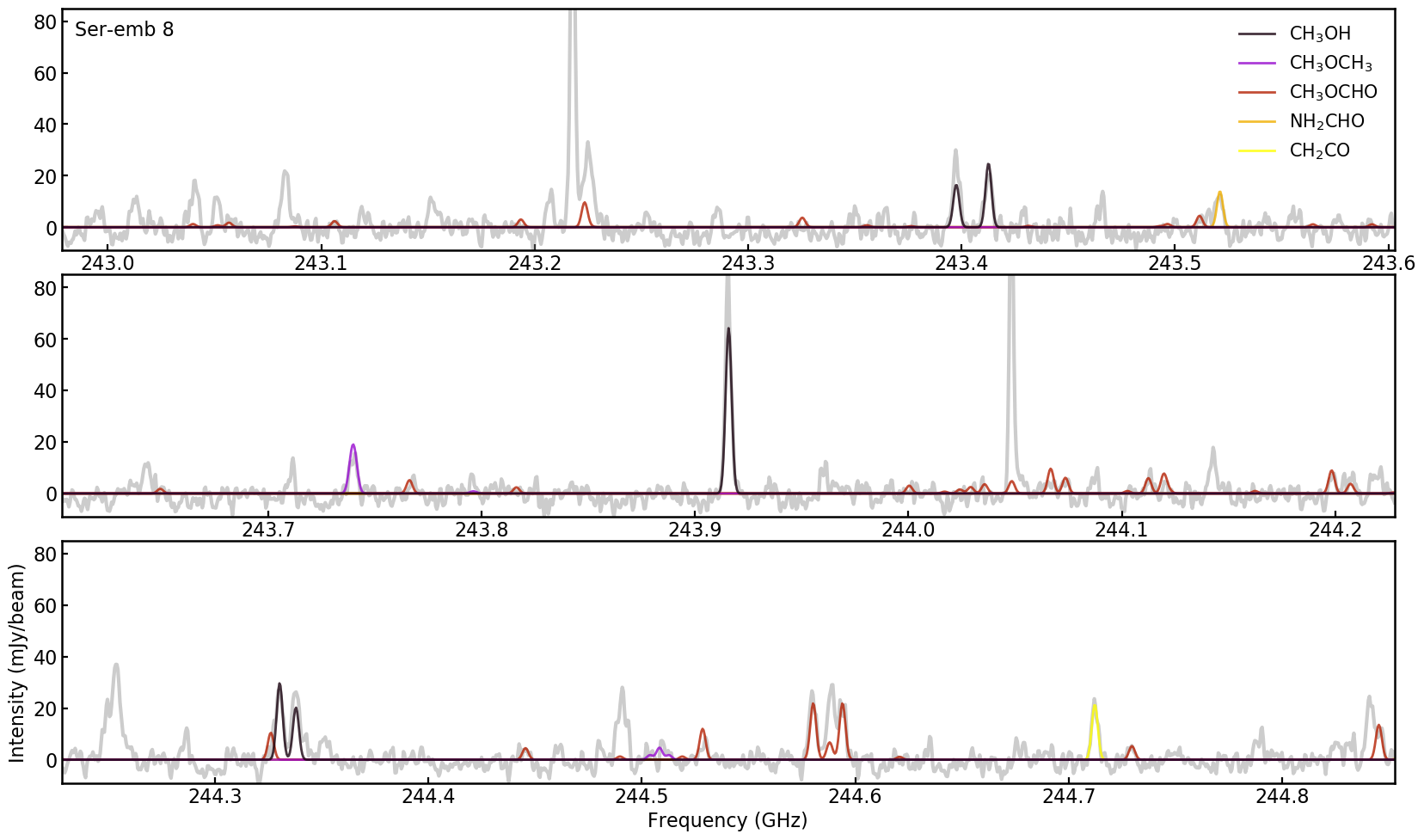}
}
\caption{Full spectrum extracted from the continuum peak pixel in Ser-emb 8 (grey line), along with synthetic spectra of the detected COMs (colored lines).  Spectra are calculated assuming the CH$_3$OH rotational temperature. \label{fig:fullspec_Ser8}}
\end{figure*}

\begin{figure*} \centering
\subfloat{
\includegraphics[clip, width=0.95\linewidth]{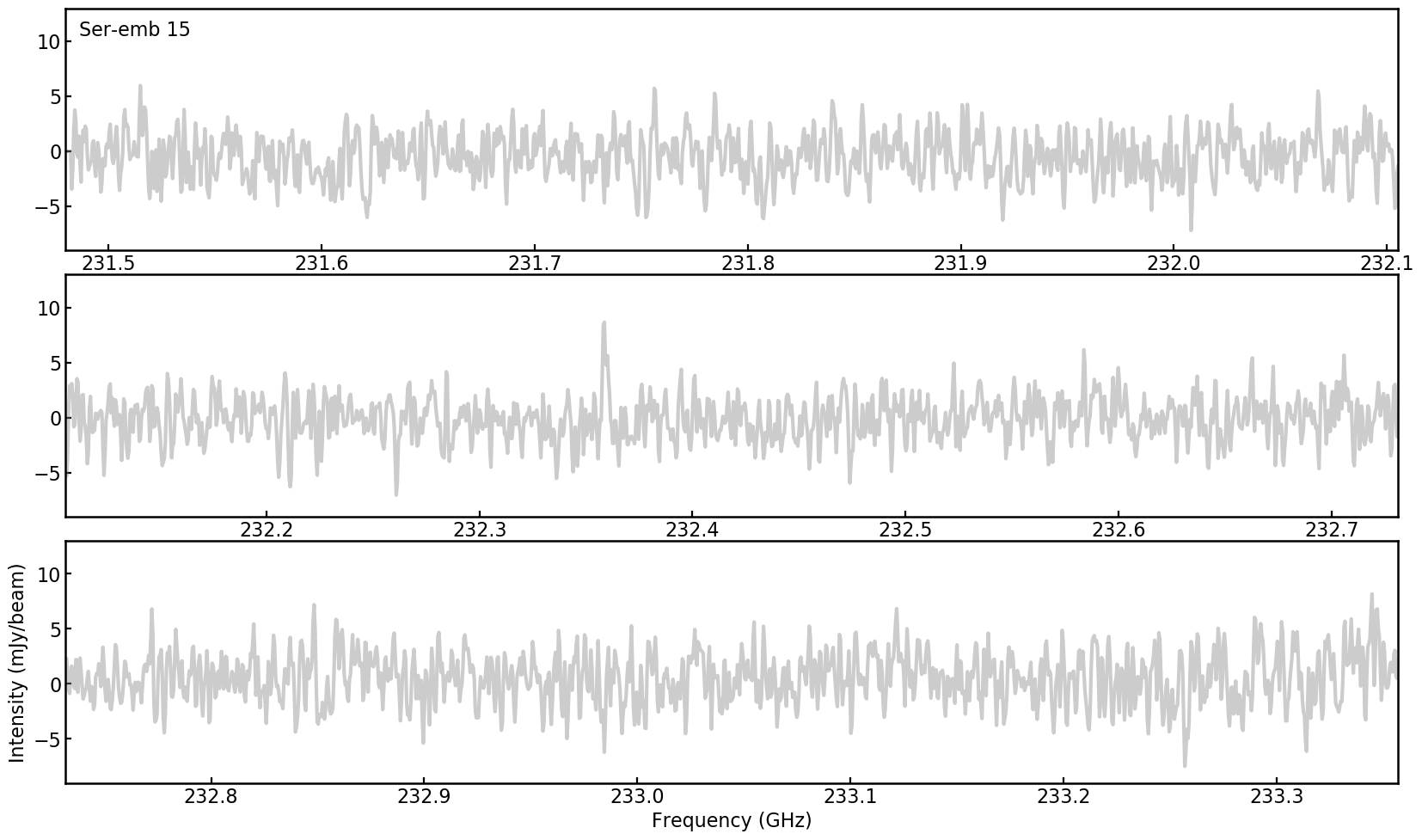}
}
\\
\subfloat{
\includegraphics[clip, width=0.95\linewidth]{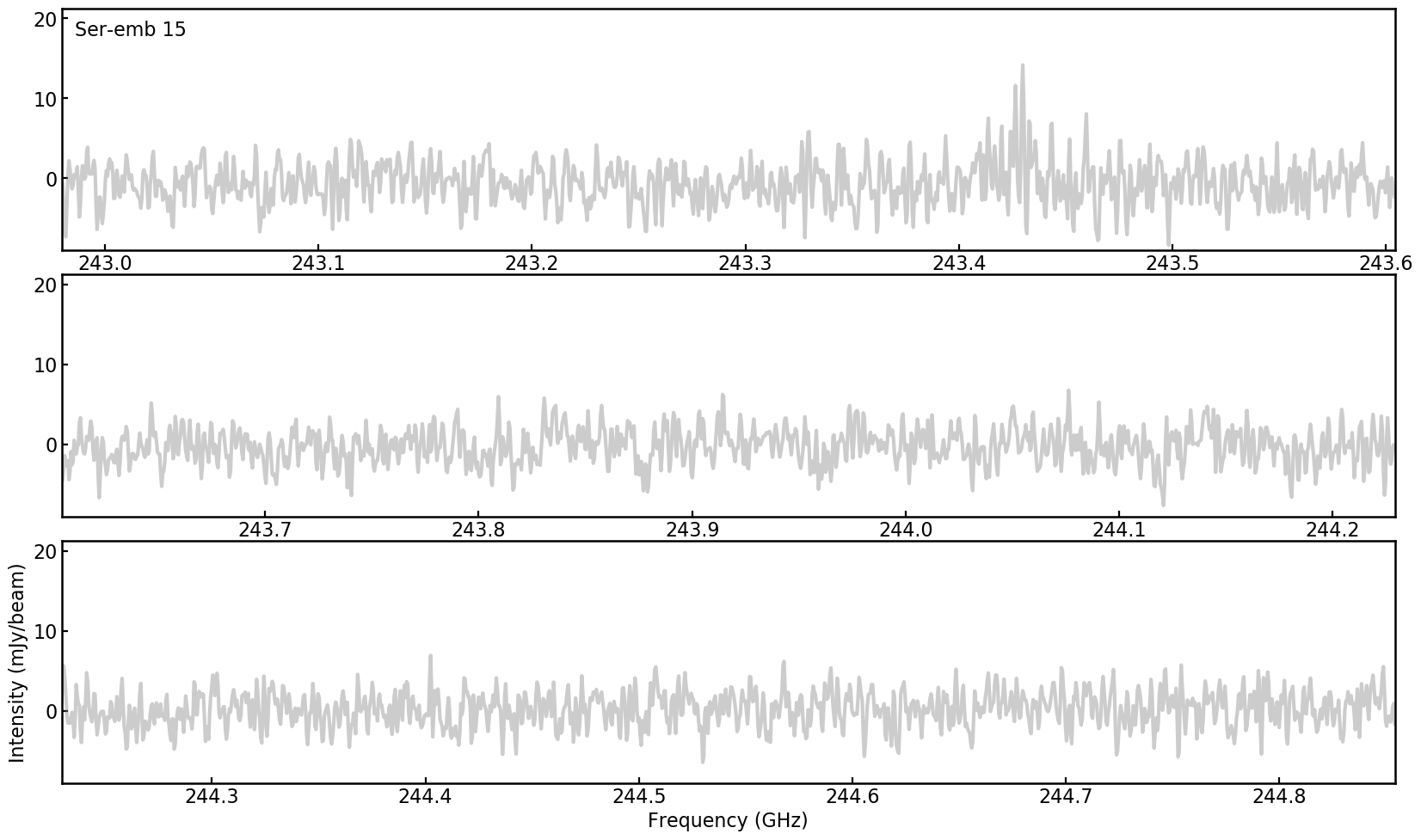}
}
\caption{Full spectrum extracted from the continuum peak pixel in Ser-emb 15 (grey line). \label{fig:fullspec_Ser15}}
\end{figure*}

\noindent Figures \ref{fig:fullspec_Ser1}--\ref{fig:fullspec_Ser15} show the full spectra extracted from the continuum peak pixel in Ser-emb 1, 7, 8, and 15, analogous to Figure \ref{fig:fullspec_Ser17}.  For Ser-emb 1 and 8 (Figures \ref{fig:fullspec_Ser1} and \ref{fig:fullspec_Ser8}) colored lines show the synthetic spectra of COMs detected in each source.

\clearpage
\bibliography{references}

\end{document}